%% file: sigir22.tex
\documentclass[sigconf]{acmart}
\AtBeginDocument{%
	\providecommand\BibTeX{{%
			\normalfont B\kern-0.5em{\scshape i\kern-0.25em b}\kern-0.8em\TeX}}}

\copyrightyear{2022}
\acmYear{2022}
\setcopyright{rightsretained}
\acmConference[SIGIR '22]{Proceedings of the 45th International ACM SIGIR Conference on Research and Development in Information Retrieval}{July 11--15, 2022}{Madrid, Spain}
\acmBooktitle{Proceedings of the 45th International ACM SIGIR Conference on Research and Development in Information Retrieval (SIGIR '22), July 11--15, 2022, Madrid, Spain}\acmDOI{10.1145/3477495.3531971}
\acmISBN{978-1-4503-8732-3/22/07}

\newcommand{\new}[1]{\textcolor{blue}{#1}}

\usepackage{tabularx}
\usepackage{multirow}
\usepackage{booktabs}
\usepackage[usestackEOL]{stackengine} 
\usepackage{cellspace}
\usepackage{makecell} 
\usepackage{caption}
\usepackage{subcaption}
\usepackage{mathtools}
\usepackage{enumitem}
\usepackage{balance}

\newcommand{\miniskip}{\vspace*{-.5\baselineskip}}
\newcommand{\shrink}{\vspace*{-.9\baselineskip}}
\usepackage{hyperref}
\begin{document}
	
	\title{Entity-aware Transformers for Entity Search}
	

	\author{Emma J. Gerritse}
	\affiliation{\institution{ Radboud University} \country{}} 
	\email{emma.gerritse@ru.nl}
	\author{Faegheh Hasibi}
	\affiliation{\institution{ Radboud University} \country{}} 
	\email{f.hasibi@cs.ru.nl}
	
	\author{Arjen P. de Vries}
	\affiliation{\institution{ Radboud University} \country{}} 
	\email{a.devries@cs.ru.nl}
	
	\renewcommand{\shortauthors}{Gerritse, et al.}
	\fancyhead{}
	\begin{abstract}
Pre-trained language models such as BERT have been a key ingredient to achieve state-of-the-art results on a variety of tasks in natural language processing and, more recently, also in information retrieval.
Recent research even claims that BERT is able to capture factual knowledge about entity relations and properties, the information that is commonly obtained from knowledge graphs.
This paper investigates the following question: 
Do BERT-based entity retrieval models benefit from additional entity information stored in knowledge graphs? 
To address this research question, we map entity embeddings into the same input space as a pre-trained BERT model and inject these entity embeddings into the BERT model. 
This entity-enriched language model is then employed on the entity retrieval task. 
We show that the entity-enriched BERT model improves effectiveness on entity-oriented queries over a regular BERT model, establishing a new state-of-the-art result for the entity retrieval task, with substantial improvements for complex natural language queries and queries requesting a list of entities with a certain property. Additionally, we show that the entity information provided by our entity-enriched model particularly helps queries related to less popular entities. 
Last, we observe empirically that the entity-enriched BERT models enable fine-tuning on limited training data, which otherwise would not be feasible due to the known instabilities of BERT in few-sample fine-tuning, thereby contributing to data-efficient training of BERT for entity search.
	\end{abstract}

	
	
\maketitle
\input{introduction}
\input{relatedwork}
\input{method}

\input{experiments}
\input{results}
\input{conclusion}


\bibliographystyle{ACM-Reference-Format}
\bibliography{sample-base}


\end{document}

%% file: introduction.tex
\section{introduction}
\label{sec:intro}
Pre-trained language models (LMs) such as BERT~ \cite{devlin:2019:bert} and its successors ~\cite{raffel:2020:exploring, yang:2019:XLNet,liu:2020:roberta} learn rich contextual information about words from large-scale unstructured corpora and have achieved intriguing results on a variety of downstream tasks in natural language processing (NLP) \cite{devlin:2019:bert, joshi:2019:bert, joshi:2019:spanbert}  and information retrieval (IR)~\cite{nogueira:2019:passage, wang:2019:multi}.
It is even shown that these language models have the capability of capturing a tremendous amount of world knowledge, including information about real-world entities otherwise found in knowledge graphs (KGs)~\cite{Petroni:2019:LMK,wang:2020:language}. For example, language models can predict masked objects in cloze sentences such as ``The native language of Mammootty is \_\_\_\_\_'' and ``The Sharon Cuneta Show was created in \_\_\_\_\_'', where each of them demonstrates a \texttt{$\langle$subject, relation, object$\rangle$} triple in a knowledge graph. 

Language models, however, fail to perform complex reasoning about entities, as they cannot capture sparse world facts~\cite{Talmor:2020:oLMpics, Petroni:2019:LMK, Jiang:2020:XMF}. In information retrieval systems, users ask complex queries about entities,  such as ``What is the second highest mountain in the world?'' or ``Most famous civic-military airports.'' Answering such queries requires leveraging rich human curated information about entities in the knowledge graphs.
To bridge the gap between LMs and KGs, recent works enhance language models with rich structured entity information from knowledge graphs, showing superior performance for knowledge-driven NLP tasks such as relation classification, entity typing, and cloze-style question answering~\cite{zhang:2019:ernie, yamada:2020:luke, poerner:2020:ebert, Jiang:2020:XMF, wang:2019:kepler,wang:2020:kadapter, peters:2019:knowledge}. Despite this success, no previous study has examined the effect of entity-enriched language models for answering entity-oriented queries in IR.

This work explores the benefit of enriching BERT-based retrieval models with auxiliary entity information from KGs for the \emph{entity retrieval} task, where users' queries are better answered with entities rather than passages or documents~\cite{Balog:2018:EOS}. It is shown that entity information improves performance of document and entity retrieval tasks~\cite{Xiong:2017:ESR, Xiong:2017:WED, Dalton:2014:EQF, Hasibi:2016:EEL, Garigliotti:2019:IET}. Yet, these studies were performed using traditional retrieval models, without utilizing LMs such as BERT.

The development of BERT-based models for entity retrieval faces a major challenge: there is limited training data for the entity retrieval task, and fine-tuning of BERT in data-constraint regimes leads to instabilities in model performance~\cite{Dodge:2020:FPL, Phang:2019:STILT, zhang:2020:revisiting}. More specifically, the official entity retrieval dataset, DBpedia-Entity v2~\cite{hasibi:2017:dbpedia}, contains 467 queries; considered a small dataset for fine-tuning of a large neural language model. When fine-tuning BERT (especially its large variant) on a target task, the training data should be large enough that the model parameters get close to the ideal setting of the target task, or otherwise it causes forgetting of what the model has already learned.

Against this background, our first and primary research question is  \emph{\textbf{RQ1:} Can an entity-enriched BERT-based retrieval model improve the performance of entity retrieval?} To address this question, we propose an entity-enriched BERT-based retrieval model, EM-BERT, where factual entity information is injected into the mono\-BERT model~\cite{nogueira:2019:passage} in the form of Wiki\-pe\-dia2Vec~\cite{yamada:2020:wikipedia2vec} entity embeddings that are transformed into the BERT's word piece vector space~\cite{poerner:2020:ebert}; see Figure~\ref{fig:bertdiagram}. The EM-BERT model is first trained with the MS MARCO passage dataset and further fine-tuned on the DB\-pe\-dia-Entity v2 collection (in a cross validation setup).  
Our experiments indicate that our EM-BERT model improves state-of-the-art entity retrieval results by 11\% with respect to NDCG@10.
We also make the intriguing finding that while fine-tuning of the plain BERT-based retrieval model (monoBERT) on DBpedia-Entity v2 is brittle and prone to degenerate performance, the same fine-tuning process using the entity-enriched BERT model (EM-BERT) results in a stable model and brings consistent and meaningful improvements. We posit that this is attributed to direct injection of entity information into the BERT model, which brings the input distribution of the model close to the output label space, thereby contributing to data-efficient training of BERT-based models for entity-oriented tasks. 

After observing the effectiveness of our EM-BERT model, we focus on understanding the whens and whys of entity-enriched BERT models for information retrieval tasks. Our next research question is: \emph{\textbf{RQ2:} When and which queries are helped by the EM-BERT model?}
We observe that EM-BERT mainly helps queries that are annotated with at least one entity; plain and entity-enriched BERT models perform on par with each other for queries without linked entities. We further examine queries with linked entities and find that EM-BERT is most helpful when entity mentions are broken into multiple word pieces by the BERT tokenizer (e.g. mention ``Tuvalu'' in Figure~\ref{fig:bertdiagram}). This indicates that directly injecting entity information into the BERT-based models is particularly important for less popular entities, which BERT does not recognize as a single word (or phrase) because they are less observed during pre-training.

In our third question, we investigate \emph{\textbf{RQ3:} Why does the EM-BERT model work and what does it learn during the fine-tuning stage?}
We approach this by comparing the embeddings produced at the final layer of the EM-BERT network for both entities and their mention tokens. We find that the learned entity representations, unlike their corresponding word piece representations, form clear clusters in the embedding space, capturing the relation between entities. We also study some examples, and show that attention weights of entity tokens are larger than other tokens for queries requesting lists of entities or queries related to less known entities.

Finally, we question \emph{\textbf{RQ4:} How does our entity-enriched BERT-based model perform on other ranking tasks?}
To address this research question, we apply the EM-BERT model to the passage ranking task, where entity information is known to be less important and a large amount of training data can be used. We observe that the entity information that is embedded in the plain BERT-based retrieval model is enough for addressing this task and auxiliary entity information from knowledge graphs does not bring significant improvements.

In summary, this work makes the following contributions:
\begin{itemize}
	\item We study the effect of enriching BERT with entity information for information retrieval tasks and propose an entity-enhanced BERT-based re-ranker EM-BERT.
	\item We establish new state-of-the-art results for the entity retrieval task on DBpedia-Entity v2 collection.
	\item We show that our entity-enriched model, unlike its equivalent model based on plain BERT, is robust against instabilities of BERT in data-constraint regimes and introduce EM-BERT as a data-efficient BERT-based model for entity retrieval.
	\item We perform thorough analysis of our EM-BERT model and add to our understanding of the model. We unfold when and for which queries this model work, what it learns, and how it works. 
\end{itemize}
The resources developed within the course of this paper are available at \url{https://github.com/informagi/EMBERT}. 

%% file: relatedwork.tex
\section{Related work}

\begin{figure*}[t]
	\centering
	\includegraphics[width=\textwidth]{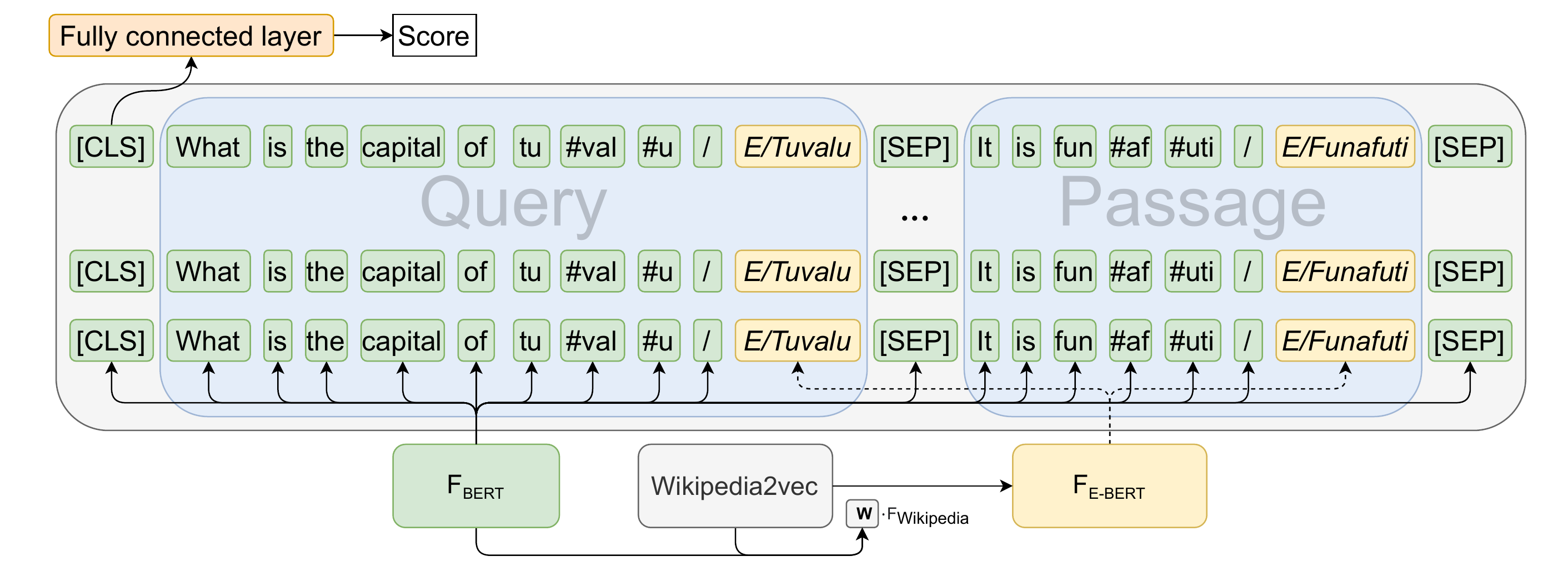}
	\caption{Illustration of the EM-BERT model. Entity annotated query and documents are tokenized and mapped to their corresponding vector representations using $F_{BERT}$ and $F_{E-BERT}$ functions.}
	\Description[]{}
	\label{fig:bertdiagram}
\end{figure*}

\paragraph{\textbf{Graph Embeddings}} \ 
After the introduction of word embeddings methods like Word2Vec \cite{mikolov:2013:distributed}, neural embedding methods became increasingly popular. Word2Vec uses a single-layer neural network to either predict a word depending on its context (Continuous Bag of Word) or the context depending on a word (Skip-Gram). The resulting hidden layer of this network captures a representation of words and can be used for other downstream tasks. After the success of these, embedding methods were expanded to capture other types of knowledge as well, for example knowledge graphs. These so-called graph embeddings aim to embed the nodes of a knowledge graph in a way that entities and their relations are mapped accordingly into the vector space. 
One of the earliest introduced methods was TransE~\cite{bordes:2013:translating}, which optimizes embeddings for head-relation-tail triples. 
\\
The graph embedding method utilized in this paper is Wiki\-pe\-dia2Vec~\cite{yamada:2020:wikipedia2vec}. This method is similar to Word2Vec, but is extended to be used on text-heavy knowledge bases like Wikipedia. Instead of solely relying on words or edges between entities, it uses a combination of words, entities, and anchor texts. Using either Skip-Gram or Continuous Bag of Words method like in Word2Vec, it employs a single layer neural network to predict neighboring words to capture word similarities, entities relations (based on neighboring entities), and word-entity relations (based on  surrounding anchor-text). Through this combination, words and entities get embedded into the same vector space.
\paragraph{\textbf{Entity Retrieval}} \
Graph embeddings have been used to assist in \emph{entity-oriented information retrieval}. 
\citet{gerritse:2020:graph} uses Wiki\-pe\-dia2Vec embeddings for entity retrieval on the DBpedia-Entity v2 collection~\cite{hasibi:2017:dbpedia}. This is done by first tagging entities in the queries by using the TAGME entity linker~\cite{ferragina:2010:TAGME}. Retrieved entities are then re-ranked by computing the cosine similarity score between the retrieved entities and the tagged entities, improving upon the state of the art for entity retrieval. KEWER~\cite{nikolaev:2020:joint} is another method for entity retrieval based on graph embeddings. It learns joint embeddings for both words and entities, combining this with entities annotated queries using SMAPH~\cite{cornolti:2016:piggyback}. KEWER also improved previous state of the art for entity retrieval.
The most recent graph-based entity ranking method, ENT Rank~\cite{dietz:2019:ENT}, utilizes entity-neighbor-text relations in a learning-to-rank model. ENT Rank incorporates a variety of entity features based on text, co-occurring entities, and entity neighbor relations,  and performs competitively on entity ranking tasks. The paper, however, uses a different version of the DBpedia-Entity v2 collection and calculates the evaluation metrics only over judged entities, making it incomparable to our and the aforementioned entity retrieval methods.

The BEIR~\cite{thakur:2021:beir} benchmark provides a number of zero-shot retrieval models for several IR databases, including the DBpedia-Entity v2 entity retrieval collection. The best performing entity retrieval model in BIER, BM25 + CE, re-ranks the top 100 entities retrieved by BM25, and then uses the MiniLM cross-encoder model \cite{wang:2020:minilm}, trained with the knowledge distillation setup provided by \citet{hofstaetter:2020:crossarchitecturekd}.



%
\paragraph{\textbf{Entity Linking}} \ 
Utilizing entity information for retrieval models requires identifying entity information from documents and queries~\cite{Hasibi:2017:ELQ, Blanco:2015:FSE, Hasibi:2015:ELQ}. This is performed by annotating text with entity linking toolkits~\cite{ferragina:2010:TAGME, cornolti:2016:piggyback, vanHulst:2020:rel, Hasibi:2017:NTE, decao:2020:autoregressive}.
In this paper, we use the REL entity linker~\cite{vanHulst:2020:rel}, which is the state-of-the-art open source entity linking toolkit. REL  detects mentions using Flair~\cite{akbik:2018:contextual} embeddings. REL performs candidate selection based on Wiki\-pe\-dia2Vec embeddings, and entity disambiguation based on latent relations between entity mentions in the text. 
\paragraph{\textbf{Transformer-based Rankers}} \ 
Bidirectional Encoder Representations from Transformers (BERT)~\cite{devlin:2019:bert} is a language representation model, pre-trained to capture bi-directional relations between words. The pre-trained BERT model can be fine-tuned for various NLP tasks. BERT can be used for binary relevance classification by feeding it a sentence A and sentence B, and using the BERT classifier token (further denoted as [CLS])  to predict relevance. This setup has been shown to be effective for question answering. 

BERT-based models are effective for \emph{information retrieval} as well, and due to their relatively low number of input tokens became especially popular for \emph{passage retrieval}. Several BERT-based methods have made enormous improvements on the previous state of the art~\cite{wang:2019:multi, nogueira:2019:passage}. MonoBERT~\cite{nogueira:2019:passage} is a point-wise re-ranker, in which the BERT model is used as a binary relevance classifier. MonoBERT, at the time of release, obtained state-of-the-art results on the MS MARCO and TREC CAR datasets. After pre-training, the ranking is done by first making an initial ranking with a baseline method like BM25, and then re-ranking a query and passage by feeding a pair of sentences A (query) and B (passage) to BERT. 


\paragraph{\textbf{Entities and Transformers}} \ 
During pre-training, BERT does not explicitly get structured data about entities. Some researchers have claimed that adding this may not be necessary since BERT captures this information implicitly~\cite{wang:2020:language, Petroni:2019:LMK}. 
Others set out to enrich BERT with entity information. 
One of the first works combining entities with transformers, ERNIE~\cite{zhang:2019:ernie}, is an enhanced language representation model, where the combination of entities and text is used to fine-tune the model. In KnowBert~\cite{peters:2019:knowledge}, contextual word embeddings are also enhanced by integrating information about entities. This is done by introducing a Knowledge Attention and Recontextualization component, which takes information in one layer of a BERT network, computes the enhanced entity information of that layer, and feeds it to the next layer. This method improves on BERT for entity-related tasks like entity linking. 

E-BERT~\cite{poerner:2020:ebert} is another method to enhance BERT with entity information. The main point which distinguishes E-BERT from the previously mentioned work is that it does not require any additional training of the network; it only requires the computation of one large transformation matrix. This allows the method to be applied to any fine-tuned transformer model. Since Wiki\-pe\-dia2Vec embeds both words and entities in the same vector space, the embeddings of words in Wiki\-pe\-dia2Vec can be used to align a mapping. 
E-BERT has been shown to improve performance on unsupervised QA tasks like LAMA, as well as downstream tasks like relation classification.

Other works combining language models and knowledge graphs include~\cite{broscheit:2019:investigating}, which enhances BERT with entities by using BERT itself for entity linking and seeing how much information of knowledge graphs is already contained in BERT. Next, BERT classifies texts to predict whether certain tokens belong to an entity. The authors show that specifically fine-tuning BERT on entity information greatly improves their scores on entity linking, thus showing that pre-trained BERT does not embed all information about entities yet, claiming that additional entity information does not help if either entities are too scarce in the data or if the task does not require entity knowledge. 

%% file: method.tex
\section{Method}
\label{sec:method}

In this section, we first provide a brief overview of knowledge graph embeddings and Wikipedia2Vec (Section~\ref{sec:method:backg}), and then describe the EM-BERT model, which is an entity-enhanced version of a BERT-based retrieval model (Section~\ref{sec:method:monoebert}), followed by a description of EM-BERT query-document input representation.



\subsection{Background}
\label{sec:method:backg}

Knowledge graph embeddings provide vector representations of entities in a knowledge graph,  projecting entity properties and relations into a continuous vector space.
These embeddings have proven effective not only for entity-related tasks such as entity linking~\cite{yamada:2016:joint}, but also for general tasks such as question answering~\cite{nikolaev:2020:joint}.
Knowledge graph embeddings can be constructed purely based on graph topology (i.e., entities and their relations)~\cite{bordes:2013:translating}, or combining graph topology with additional textual information such as the entity descriptions~\cite{yamada:2016:joint}. 
In this paper, we use Wikipedia2Vec embeddings~\cite{yamada:2020:wikipedia2vec}, which extends the skip-gram model~\cite{mikolov:2013:distributed} with entity relations and descriptions from Wikipedia and maps words and entities into a shared vector space. Following~\cite{poerner:2020:ebert}, we transform Wikipedia2Vec entity embeddings into the BERT word piece embedding space and use them as input to the BERT model, similar to a BERT word piece vector.

\begin{table*}[t]
	\caption{Statistics about document length and number of linked entities in queries and documents of the used datasets, with mean and standard deviation. Document and query length information is computed without adding entity information.}
	\label{tab:dataset_info}
	\begin{tabular}{ l || l l l l }
		\hline
		Dataset & size& avg length & avg entities &  total linked \\
		\Xhline{2pt}
		MS MARCO dev queries & 6980 & $6.95\pm 2.54$& $0.34 \pm 0.57 $  & 2392 \\ 
		MS MARCO passages & 8841823 & $58.27 \pm 22.94$ &$2.55 \pm 3.22$ & 22532200 \\
		DBpedia-Entity v2 queries & 467 &$ 6.55 \pm 2.88$ & $0.82 \pm 0.64$ & 381 \\
		DBpedia-Entity v2 short abstracts & 389622 & $55.93 \pm 25.95$ & $6.88 \pm 3.93$ & 2678843 \\
		\hline
		
	\end{tabular}
\end{table*}

Formally, Wikipedia2Vec embeds a list of words  $\mathbb{L}_{word}$ and a list of entities $\mathbb{L}_{ent}$ into a vector space $\mathbb{R}^{d_{\mathit{Wikipedia}}}$, where $d_{\mathit{Wikipedia}}$ is dimensions of the Wikipedia embeddings. 
The embeddings are trained by fitting a one layer neural network with the loss:
\begin{equation*}
	\mathcal{L} = \mathcal{L}_w + \mathcal{L}_e + \mathcal{L}_a,
\end{equation*}
which consist of three cross-entropy loss functions:
\begin{itemize}
	\item[$\mathcal{L}_w$:] predicting context words of a given word.
	\item[$\mathcal{L}_e$:\ ]  predicting neighboring entities of a given entity.
	\item[$\mathcal{L}_a$:\ ]  predicting anchor text of an entity to combine words and entities in a single embedding space.
\end{itemize} 

The weight layer of this neural network results in the function  $F_{\mathit{Wikipedia}}\ :\mathbb{L}_{word}\cup\mathbb{L}_{ent} \rightarrow \mathbb{R}^{d_{\mathit{Wikipedia}}}$, which embeds both words and entities in the Wikipedia vector space. BERT, on the other hand, tokenizes text into the word piece dictionary $\mathbb{L}_{wordpiece}$. We use the  lookup function $F_{BERT}:\mathbb{L}_{wordpiece} \rightarrow \mathbb{R}^{d_{BERT}}$ to transform word pieces into the $d_{BERT}$-dimensional native BERT word piece vector space, the resulting vector being the input for the BERT model.

\subsection{EM-BERT}
\label{sec:method:monoebert}
Our proposed model, referred to as EM-BERT, incorporates entity embeddings into a point-wise document ranking approach. In its essence, EM-BERT combines \underline{E}-BERT~\cite{poerner:2020:ebert} with \underline{m}onoBERT~\cite{nogueira:2019:passage} and predicts a retrieval score for a query-document pair, each represented by a sequence of words and entities. The model takes as input the concatenation of the query tokens $t_{q_1}, ..., t_{q_n}$ and the document tokens $t_{d_1}, ..., t_{d_m}$, where each token $t_i$ is either a BERT native word piece token or an entity token.


\paragraph{\textbf{Entity-enriched BERT}} \ 
The off-the-shelf BERT model only accepts its native word pieces. To incorporate entity embeddings into BERT, we need to align entity vectors with word piece vectors. Following~\cite{poerner:2020:ebert}, this is done by a linear transformation of entity vectors to BERT-like vectors.
Since $\mathbb{L}_{wordpiece}$ does not contain any entities, the Wikipedia2Vec word dictionary $\mathbb{L}_{Word}$ is used to obtain the linear transformation $\mathbf{W} \in \mathbb{R}^{ d_{BERT}\times d_{\mathit{Wikipedia}}}$, 
learned from the Wikipedia2Vec word vector space $F_{\mathit{Wikipedia}}[\mathbb{L}_{Word}]$ and BERT word piece space $F_{BERT}[\mathbb{L}_{wordpiece}]$:

%
\begin{equation} 
	\label{eq:ebert}
	\mathbf{W} = argmin_{W} \sum_{\mathclap{x \in\mathbb{L}_{word}\cap \mathbb{L}_{wordpiece}}} || W\cdot F_{\mathit{Wikipedia}}(x) - F_{BERT}(x)||^2_2.
\end{equation}
Here, the intersection between $\mathbb{L}_{word}$ and $ \mathbb{L}_{wordpiece}$ is taken to ensure that the selected words have embeddings in both Wikipedia and BERT vector spaces. 
The equation computes matrix $\mathbf{W}$, in which the distance between 
$\mathbf{W}\cdot F_{\mathit{Wikipedia}}(x)$ and $F_{BERT}(x)$ is minimal for all $x$. Using $\mathbf{W}$, we can then construct the function $F_{E-BERT}$,
which maps both entities and word tokens to the BERT input vector space:
%
%
\begin{equation} 
	\label{eq:embert}
	F_{E-BERT}(a) = 
	\begin{cases}
		\mathbf{W}\cdot F_{\mathit{Wikipedia}}(a),& a \in  \mathbb{L}_{ent}\\
		F_{\mathit{wordpiece}}(a),&\text{otherwise}.
	\end{cases}
\end{equation}
The aligned entity vectors are then fed into BERT when an entity is mentioned in the input text. The entities are obtained by annotating the text with an entity; e.g., given the text ``Who produced films starring Natalie Portman", the annotated sentence becomes  ``Who produced films starring Natalie Portman ENTITY/\-Natalie\-Portman". This text is then tokenized as ``\texttt{who produced films starring natalie port \#\#man \textbf{/ENTITY/Natalie\_Portman}}," where the entity ID (in bold face) is a single token, embedded by $F_{E-BERT}$ (Eq.~\ref{eq:embert}). We note that the transformation matrix $\mathbf{W}$ can be fit into any pre-trained or fine-tuned BERT model, thus making E-BERT embeddings usable for every available BERT model. 

\paragraph{\textbf{Retrieval}} \
The retrieval phase is based on monoBERT~\cite{nogueira:2019:multi}, which is a multi-stage ranking method with BERT and has been shown to achieve competitive results on MS MARCO passage retrieval~\cite{nguyen:2016:msmarco} and TREC-CAsT~\cite{dalton:2020:trec}. The simplicity of this model, coupled with its high effectiveness enables us to discern the effect of entity information on BERT-based retrieval models and obtain an understanding of the whys and whens of entity-enriched BERT models for information retrieval tasks. 

In the multi-stage ranking method, for a query $q$, documents are ranked with an initial ranking method; e.g., BM25. The top-$k$ documents, denoted as $D_K = \{d_1, d_2, \dots,  d_k\}$, are then passed to a second ranker for re-ranking; here a BERT-based retrieval model. Every query-document pair $\langle q, d_i\in D_K \rangle$ is passed to BERT as two sentences $A$ and $B$, with the separation token [SEP] in between. The BERT [CLS] classification vector is used as input to a single layer neural network to obtain the probability $s_i$ of the document $d_i$ being relevant to the query $q$. To train this model, the cross-entropy loss is used:
\begin{equation}
	L = - \sum\limits_{j\in J_{pos}}log(s_j) - \sum\limits_{j\in J_{neg}}log(1-s_j),
\end{equation}
where $J_{pos}$ and $J_{neg}$ are indexes of relevant and non-relevant documents for all queries.

Putting all the pieces together, the training process of the EM-BERT model is summarized as follows. First, the transformation matrix $\mathbf{W}$ is trained using Wikipedia2Vec embeddings and BERT word pieces. All queries and documents are annotated with an entity linker, and the tagged queries and documents are tokenized and mapped to the corresponding vector representations using function $F_{E-BERT}$ . This input is fed into the EM-BERT model and the model is first fine-tuned on MS MARCO passages and then on the entity retrieval collection.

\paragraph{\textbf{Query-Document Representation}}
In the EM-BERT model (similar to monoBERT), queries are truncated to have a maximum token length of 64, and the combination of the query, passage, and separation tokens are truncated to have a maximum length of 512 tokens.
Following~\cite{poerner:2020:ebert}, we concatenate mentions and entities, separated by the token `/'; i.e.,  ``\texttt{mention + `/' + EntityID}.'' The final representation of query-document pairs is ``[CLS] + Query + [SEP] + Document + [SEP];'' see  Figure \ref{fig:bertdiagram}.

%% file: experiments.tex
\section{Experimental setup}
\label{sec:exp}

\begin{table*}[t]
	\shrink
	\centering
	\caption{Results on DBpedia-Entity v2 collection. Superscripts 1/2/3/4 denote statistically significant differences (better or worse) compared to base method/monoBERT (1st)/monoBERT/EM-BERT (1st), respectively. The highest value per column is marked with underline and in bold. The highest value per column and per block is marked with underline.}
	\label{tab:dbpedia}
	\begin{tabular}{p{0.17\linewidth}@{~}||  ll |  ll|  ll | ll||  ll}
		
		
		\hline

		\multirow{2}{*}{NDCG}
		& \multicolumn{2}{Sc |}{\textbf{SemSearch}} & \multicolumn{2}{c |}{\textbf{INEX-LD}} 
		& \multicolumn{2}{c |}{\textbf{ListSearch}} & \multicolumn{2}{c ||}{\textbf{QALD-2}} 
		& \multicolumn{2}{c }{\textbf{Total}}\\
		&
		\multicolumn{1}{l}{@10} & \multicolumn{1}{l|}{@100} & 
		\multicolumn{1}{l}{@10} & \multicolumn{1}{l|}{@100} &
		\multicolumn{1}{l}{@10} & \multicolumn{1}{l|}{@100} &
		\multicolumn{1}{l}{@10} & \multicolumn{1}{l||}{@100} &
		\multicolumn{1}{l}{@10} & \multicolumn{1}{l}{@100}\\

		\Xhline{2pt}
		\small{ESIM$_{cg}$ \cite{gerritse:2020:graph}} &0.417 & 0.478 & 0.217 & 0.286 & 0.211 & 0.302 & 0.212 & 0.282 & 0.262 & 0.335\\
	\small{KEWER \cite{nikolaev:2020:joint}} & - & - & - & - & - & - & - &- & 0.270& 0.310 \\
		\small{BLP-TransE \cite{daza:2021:inductive}} &  0.631 & 0.723 & 0.446 & 0.546 & 0.442& 0.540& 0.401& 0.482& 0.472& 0.562 \\
		
		\small{BM25F+KEWER \cite{nikolaev:2020:joint}} & 0.661 & 0.733 & 0.468 & 0.530 & 0.440 & 0.521 & 0.386 & 0.474 & 0.483 & 0.560 \\

		\hline
		BM25 & 0.425 & 0.523 & 0.298 & 0.330 & 0.274 & 0.322 & 0.192 & 0.243 & 0.291 & 0.349 \\ 
		+monoBERT (1st) & $0.588^{1}$ & $0.655^{1}$ & $0.406^{1}$ & $\underline{0.459}^{1}$ & $0.422^{1}$ & $0.458^{1}$ & $0.350^{1}$ & $0.406^{1}$ & $0.437^{1}$ & $0.490^{1}$ \\
		+monoBERT  & $0.577^{1}$ & $0.640^{12}$ & $\underline{0.409}^{1}$ & $0.452^{1}$ & $0.423^{1}$ & $0.449^{12}$ & $0.347^{1} $& $0.392^{12}$ & $0.435^{1}$ & $0.479^{12}$\\
		+EM-BERT (1st)  & $0.593^{1}$ & $0.667^{13}$ & $0.395^{1}$ & $0.448^{1}$ & $0.436^{1}$ & $0.465^{1}$ & $0.351^{1}$ & $0.403^{1}$ & $0.440^{1}$ & $0.492^{13}$ \\ 
		+EM-BERT & $\underline{0.612}^{1}$ & $\underline{0.672}^{1}$& $0.392^{1}$ & $0.434^{12}$ & $\underline{0.478}^{1234}$ & $\underline{0.469}^{13}$ & $\underline{0.375}^{1234}$ & $\underline{0.418}^{134}$ & $\underline{0.461}^{1234}$ & $\underline{0.495}^{13}$ \\ 		\hline

		BM25F-CA \cite{hasibi:2017:dbpedia} & $\underline{0.628}^{}$ & $\underline{0.720}^{}$ & $\underline{0.439}^{}$ & $\underline{0.530}^{}$ & $0.425^{}$ & $0.511^{}$ & $0.369^{}$ & $0.461^{}$ & $0.461^{}$ & $0.551^{}$ \\	
		+monoBERT (1st) & $ 0.605^{} $ & $ 0.682^{1} $ & $ 0.427^{} $ & $ 0.501^{} $ & $ 0.440^{} $ & $ 0.510^{} $ & $ 0.405^{} $ & $ 0.490^{} $ & $ 0.467^{} $ & $ 0.544^{} $\\
		+monoBERT  & $ 0.591^{2} $ & $ 0.663^{12} $ & $ 0.434^{} $ & $ 0.489 ^{12} $& $ 0.438^{} $ & $ 0.496^{2} $ & $ 0.401^{} $ & $ 0.467^{2} $ & $ 0.463^{} $ & $ 0.526^{12} $\\
		+EM-BERT (1st) & $ 0.609^{} $ & $ 0.695^{3} $ & $ 0.418^{} $ & $ 0.490^{1} $ & $ 0.452^{} $ & $ 0.518^{3} $ & $ 0.398^{} $ & $ 0.483^{} $ & $ 0.466^{} $ & $ 0.544 ^{3} $\\
		+EM-BERT  & $0.621^{} $ & $ 0.695^{} $ & $ 0.399^{} $ & $ 0.469^{12} $ & $ \underline{0.492}^{1234} $ & $ \underline{0.535}^{23} $ & $ \underline{0.435}^{1234} $ & $ \underline{0.512}^{1234} $ & $ \underline{0.486}^{1234} $ & $ \underline{0.553}^{3} $\\
		\hline

		\small{GEEER} \cite{gerritse:2020:graph} &  $0.660^{} $ & $ 0.736^{} $ & $ 0.466^{} $ & $ 0.552^{} $ & $ 0.452^{} $ & $ 0.535^{} $ & $ 0.390^{} $ & $ 0.483^{} $ & $ 0.487^{} $ & $ 0.572^{} $\\
		+monoBERT (1st)   & $ 0.643^{} $ & $ 0.733^{} $ & $ \mathbf{\underline{0.486}^{}} $ & $ \mathbf{\underline{0.564}^{}} $ & $ 0.493^{1} $ & $ 0.554^{1} $ & $ 0.449^{1} $ & $ 0.525^{1} $ & $ 0.515^{1} $ & $ 0.591 ^{1} $\\
		+monoBERT  & $ 0.633^{2} $ & $ 0.725^{2} $ & $ 0.481^{} $ & $ 0.563^{} $ & $ 0.492^{1} $ & $ 0.553^{1} $ & $ 0.439^{12} $ & $ 0.519^{12} $ & $ 0.508^{12} $ & $ 0.586^{12} $ \\
		+EM-BERT (1st)  & $ {{0.650}^{}} $ & $ 0.743^{} $ & $ 0.483^{} $ & $ 0.561^{} $ & $ 0.504^{1} $ & $ 0.562^{1} $ & $ 0.444^{1} $ & $ 0.522^{1} $ & $ 0.517^{13} $ & $ 0.594^{13} $ \\
		+EM-BERT & $ \mathbf{\underline{{0.664^{}}}} $ & $ \mathbf{\underline{0.744}}^{} $ & $ 0.479^{} $ & $ 0.561^{} $ & $ \mathbf{\underline{0.544}}^{1234} $ & $ \mathbf{\underline{0.579}}^{1234} $ & $ \mathbf{\underline{0.483}}^{1234} $ & $ \mathbf{\underline{0.543}}^{1234} $ & $ \mathbf{\underline{0.541}}^{1234} $ & $ \mathbf{\underline{0.604}}^{1234} $  \\
		\hline
	\end{tabular}
\end{table*}

To address our research question, we compare the EM-BERT model with several baselines on the entity retrieval task and analyze the results.  Due to limited training data for entity retrieval, we perform two-stage fine-tuning: we first fine-tune our models on the annotated MS MARCO passage dataset, and then continue fine-tuning on DBpedia-Entity v2 collection. In the following, we describe our experimental setup. 

\subsection{Entity Linking and Embeddings}
We use the REL entity linking toolkit ~\cite{vanHulst:2020:rel} to annotate texts with entities. REL is known to have high precision compared to other entity linking toolkits such as GENRE~\cite{decao:2020:autoregressive} and TAGME~\cite{ferragina:2010:TAGME}, ensuring that limited noise (incorrectly annotated entities) is added to the model. REL incorporates Wiki\-pe\-dia2Vec embeddings trained on the Wikipedia link graph. To avoid a mismatch between annotated entities and Wiki\-pe\-dia2Vec entity embeddings~\cite{gerritse:2020:graph, dietz:2019:ENT}, we follow the REL Wiki\-pe\-dia2Vec training procedure, using the Wikipedia dump of 2019-07-01, setting \textit{min-entity-count} parameter to zero, with a dimensionality of 500.


\subsection{First Stage Fine-tuning}
\label{sec:exp:passage}

\paragraph{\textbf{Collection}} 
For the first stage fine-tuning of EM-BERT, we use the MS MARCO passage ranking collection, consisting of 8,841,823 passages, extracted from web documents retrieved by Bing.
MS MARCO provides a training set with approximately 800k queries, a development set with 6980 queries with public relevance judgments, and an evaluation set with 6837 queries and private relevance judgments. 

Although MS MARCO passage ranking is not an entity-related task, REL annotates 29\% of queries in the development sets with at least one entity; e.g., ``what were the \emph{brothers grimm} names.'' The statistics of this collection (with REL annotations) are reported in Table~\ref{tab:dataset_info}. It shows that on average every query and document is annotated with 0.34 and 2.55 entities respectively, indicating that a fair amount of entities is present in this dataset. 
%
\paragraph{\textbf{Training}} \ 
We fine-tune our EM-BERT model on the officially provided MS MARCO training set. 
This set provides triples of a query, a positive example, and a negative example, making the ratio of positive-negative documents equal during training. We refer to these triples as ``sample.''
Starting from the monoBERT model provided by HuggingFace,\footnote{\url{https://huggingface.co/castorini/monobert-large-msmarco}} we fine-tune EM-BERT with randomly selected 300k samples (cf.~\S\ref{sec:method:monoebert}); fine-tuning with more samples, up until 900k, did not give any further improvements on the development set.
We use an accumulated batch size of 64, a learning rate of $10^{-6}$, 40k warm-up steps, and AdamW as optimizer.
Fine-tuning for 300k samples took around 34 hours on a single GeForce RTX 3090 with 24 GB memory.
%

\subsection{Second Stage Fine-tuning} 
\label{sec:exp:er}
\paragraph{\textbf{Collection}} \ 
For the second stage fine-tuning, we use the standard entity retrieval collection, DB\-pe\-dia-Entity v2~\cite{hasibi:2017:dbpedia}. The collection is based on DBpedia 2015-10 and consists of four different categories of queries: (i) \emph{SemSearch}: named entity queries looking for particular entities; e.g., ``mario bros'', (ii) \emph{INEX-LD}: keyword-style entity-oriented queries; e.g., ``bicycle benefits health'', (iii) \emph{ListSearch}: queries asking for a list of entities; e.g., ``Hybrid cars sold in Europe'', and (iv) \emph{QALD-2}: natural language questions about entities; e.g., ``Who produces Orangina?'' Relevance judgements in this data\-set are graded: Highly relevant (2), Relevant (1), and irrelevant (0). 

Documents in the DBpedia-Entity v2 collection are entities, each represented by an abstract and other metadata (stored in RDF format). These entity abstracts provide concise descriptions about entities and are comparable in length to MS MARCO passages. Both queries and entity abstracts are highly entity-oriented, reflected by the average number of entity annotations for queries and abstracts in Table~\ref{tab:dataset_info}.


\paragraph{\textbf{Training}} \ 
We consider entity abstracts as entity descriptions and use them as documents in monoBERT and EM-BERT. Since the DBpedia-Entity v2 collection provides a limited amount of training data (only 467 queries), we take the fine-tuned model on  the MS MARCO passage dataset (Section~\ref{sec:exp:passage}) and continue fine-tuning for both monoBERT and EM-BERT models using the provided folds for 5-fold cross validation, each fold containing ~48k query-entity pairs with approximately 34\% relevant and 66\% non-relevant entities.
In this process, both models are fine-tuned for one epoch using all training data of each fold, resulting in 5 different models each. The fine-tuned models are then used for ranking the test queries in the corresponding folds. Note that in this fine-tuning process, no hyper-parameter optimization is performed.  
In this process, both models are fine-tuned for one epoch using the training data of each fold, resulting in 5 different models each. The fine-tuned models are used for ranking the test queries in the corresponding folds. Note that in this fine-tuning process, no hyper-parameter optimization is performed.  
For both models an aggregated batch size of 64, learning-rate of $10^{-6}$, and AdamW optimizer is used. Warmup step is changed to 4000, following the 10\% warmup step rule.

\subsection{Baselines}
\label{sec:exp:baselines}
We compare EM-BERT with the state-of-the-art entity retrieval methods using term-matching, neural, and BERT-based approaches. The baselines are:

\begin{figure*}[t]
	\begin{minipage}[c]{0.48\textwidth}
		\includegraphics[width=0.99\columnwidth]{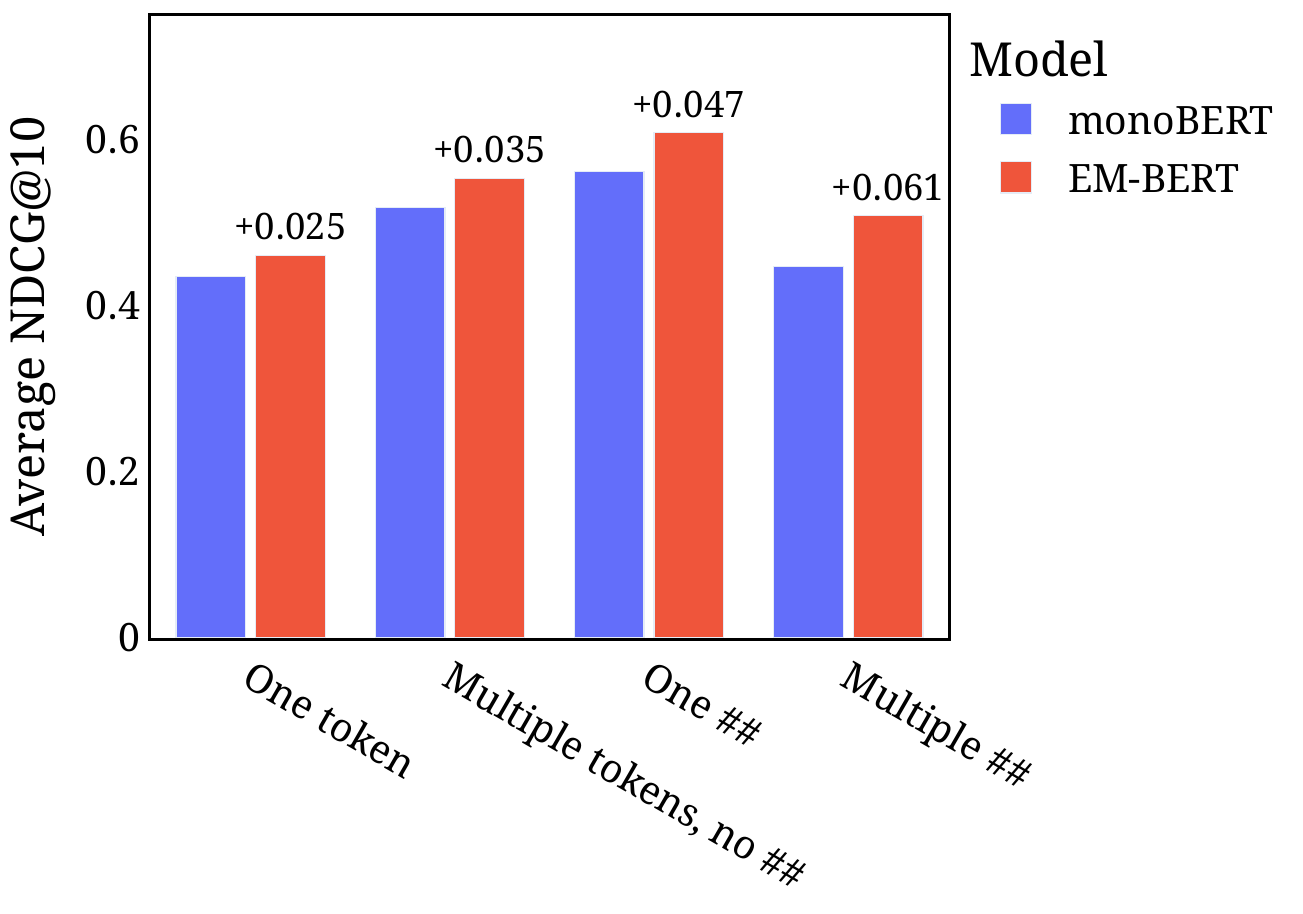}
		\miniskip
		\caption{Comparison of re-ranked BM25F-CA run with monoBERT and EM-BERT for different mention tokenization categories.}
		\Description[]{}
		\label{fig:tokenization}
	\end{minipage}%
	\hfill
	\begin{minipage}[c]{0.48\textwidth}
		\shrink
		\includegraphics[width=0.99\columnwidth]{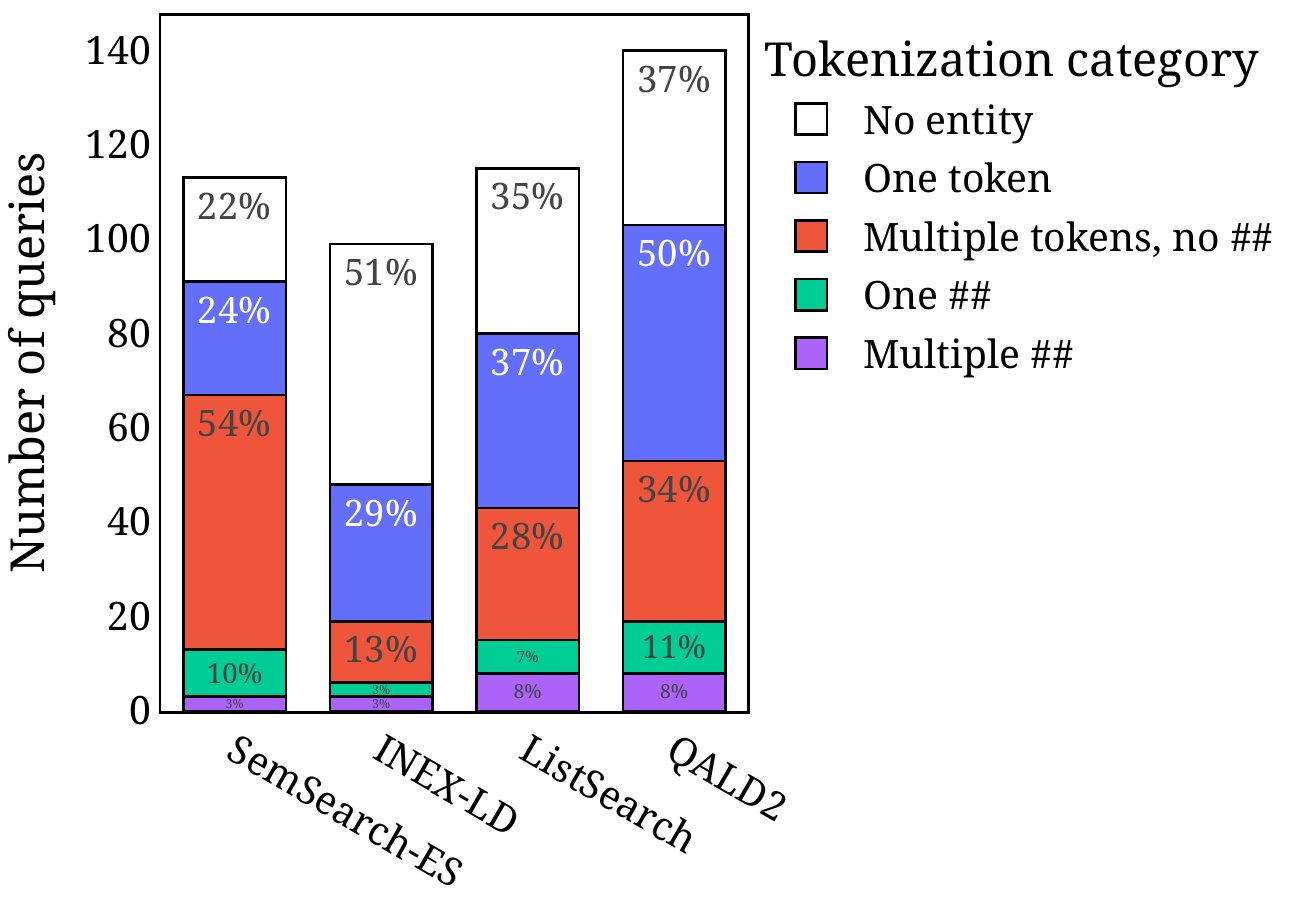}
		\miniskip
		\caption{Mention tokenization categories for the for types of queries in DBPedia-Entity v2 collection.}
	\Description[]{}
	\label{fig:heatmap}
\end{minipage}
\end{figure*}

\begin{itemize}[leftmargin=0.9em]

\item []\textbf{BM25}: The BM25 run, based on the short abstracts of entities, with parameters $k = 0.9$ and $b = 0.4$. Similar to monoBERT and EM-BERT, this run uses only abstract information of entities (unlike other methods that use reach representation of entities). 

\item []\textbf{BM25F-CA~\cite{hasibi:2017:dbpedia}}: The fielded variant of BM25, obtaining scores  from 5-field representation of entities. Field weights are computed using Coordinate Ascent on each train fold, and then scored on the corresponding test fold. This is the best non-neural run and also the best baseline reported in~\cite{hasibi:2017:dbpedia}.

\item []\textbf{ESIM$_{cg}$~\cite{gerritse:2020:graph}}: ESIM$_{cg}$ ranks entities based on the aggregated cosine similarity scores between the Wikipedia2Vec embedding of each linked entity in the query and the target entity. Entities in this method are linked using TAGME~\cite{ferragina:2010:TAGME}. The method re-ranks the BM25F-CA run.
\item[]\textbf{GEEER~\cite{gerritse:2020:graph}}: GEEER  linearly combines the BM25F-CA score and entity-query similarities based on ESIM$_{cg}$. The weights of this linear combination are computed on each train fold using Coordinate Ascent, and scored accordingly on each test fold. Similar to ESIM$_{cg}$, GEEER re-ranks the BM25F-CA run.
\item []\textbf{KEWER \cite{nikolaev:2020:joint}}: KEWER uses an embedding method with joint word and entity embeddings. A query-entity score is computed on all query terms by the weighted sum of the similarity between the target entity embedding and the term, re-ranking the BM25F-CA run.
\item []\textbf{BM25F+KEWER \cite{nikolaev:2020:joint}}:  The model linearly combines the KEWER score as listed above with BM25F-CA and re-ranks the BM25F-CA run. 
\item []\textbf{BLP-TransE \cite{daza:2021:inductive}}: BLP-TransE is a BERT-based method, combining a BERT cross-encoder architecture with TransE graph embeddings. It encodes queries and entities, uses the similarity between these encodings as query score, and then re-ranks and linearly combines the BM25F-CA scores.  

\item []\textbf{monoBERT}:
Our Pygaggle implementation of the monoBERT model. This implementation achieves slightly better performance compared to results reported in~\cite{nogueira:2019:multi}: MRR@10 of 0.379 vs 0.372 on the MS MARCO development set. Following the multi-stage re-ranking method~\cite{nogueira:2019:multi}, we re-rank three entity retrieval runs using monoBERT: BM25, BM25F-CA, and GEEER (the best baseline run). The re-ranking is performed on the top-1000 entities of the BM25 and BM25F-CA, and top-100 entities of the GEEER run (as GEEER re-ranks the BM25F-CA run).

\end{itemize}
\begin{table}[t]
	\caption{Comparison of results for queries without and with at least one linked entity. 
		Superscripts 1/2/3 show statistically significant differences compared to BM25F-CA/monoBERT (1st)/monoBERT, respectively.}
	\label{tab:difference_annotated_dbpedia}
	\begin{tabular}{ l ||@{~}l  l@{~}}
		\hline
		DBpedia-Entity v2 & NDCG@10 (>1 en) & NDCG@10 (no-en) \\
		\Xhline{2pt}
		BM25F-CA & 0.481 & 0.414  \\
		+monoBERT (1st) & 0.487 & 0.422  \\
		+monoBERT & 0.482 & 0.420  \\
		+EM-BERT &$0.516^{123}$ & 0.420 \\
		\hline
	\end{tabular}
	\shrink
\end{table}

\shrink
\subsection{Evaluation Metrics}
For evaluation of entity retrieval, we use the commonly reported metric in~\cite{hasibi:2017:dbpedia}; the Normalized Discounted Cumulative Gain (NDCG) at ranks 10 and 100. Statistical significant differences of NDCG@10 and NDCG@100 values are determined using the two-tailed paired t-test with p-value $<0.05$.


%% file: results.tex
\miniskip
\section{Results and Analysis}
\label{sec:res}

In the following, we evaluate our entity-enriched BERT-based retrieval model and answer our four research questions (Sections~\ref{sec:res:er}-\ref{sec:res:other} ) listed in Section~\ref{sec:intro}.

\subsection{Entity Retrieval Results}
\label{sec:res:er}
In this section, we answer our first research question: \emph{\textbf{RQ1:} Can an entity-enriched BERT-based retrieval model improve the performance of entity retrieval?} 

\noindent
We compare retrieval performance of EM-BERT with a variety of models on the
DBpedia-Entity v2 collection in Table~\ref{tab:dbpedia}. The baseline (cf. Section~\ref{sec:exp:baselines}) results are presented in the top part of the table.
The next three compartments in the Table summarise results of re-ranking BM25, BM25F-CA, and GEEER runs with the following models:
\begin{itemize}[leftmargin=2em]
	\item \textbf{monoBERT (1st)}: The monoBERT model after first stage fine-tuning; cf. Section~\ref{sec:exp:baselines}.
	\item \textbf{monoBERT}: The monoBERT model after second stage fine-tuning on DBpedia-Entity v2.  
	\item \textbf{EM-BERT (1st)}: The EM-BERT model fine-tuned on the MS MARCO passage collection. Re-ranking setup is similar to the monoBERT runs; cf. Section~\ref{sec:exp:baselines}. 
	\item \textbf{EM-BERT}: The EM-BERT model after second stage fine-tuning. Setup is similar to the previous run.
\end{itemize}

As shown in Table~\ref{tab:dbpedia}, re-ranking of the GEEER run with EM-BERT markedly outperforms all baselines and establishes a new state-of-the-art result on the DBpedia-Entity v2 collection, with 11\% improvements over the best run. 
When re-ranking the BM25 and BM25F-CA runs, we observe the same trend that EM-BERT outperforms all the corresponding BERT-based baselines. 
Comparing different query categories, EM-BERT improves over all runs by a large margin for ListSearch and most importantly for QALD-2 queries; improvements over GEEER are 20\% and 24\% with respect to NDCG@10 for ListSearch and QALD-2 queries, respectively. QALD queries are complex natural language queries that are hard to answer for most neural and non-neural models, reflected by the lowest NDCG scores on all baseline models.

%

Table~\ref{tab:dbpedia} also shows that monoBERT performance is decreased when fine-tuned with a limited amount of training data. The drop in the performance of monoBERT is expected due to the known instabilities of BERT with few-sample training~\cite{zhang:2020:revisiting}, which can cause forgetting of what the model has already learned. The striking observation, however, is that EM-BERT results are improved when trained on the same data and using the same procedure. This indicates that entity-enriched BERT models enable fine-tuning on limited training data, thereby contributing to data-efficient training of BERT for entity-oriented tasks.

Based on these results, we can positively answer our first research question: \emph{our entity-enriched BERT-based model significantly improves state-of-the-art results on entity retrieval, and more interestingly, it is robust against instabilities of BERT when fine-tuned with limited training data.}

\begin{figure*}[t]
	\miniskip
	\centering
	\begin{subfigure}[t]{0.33\textwidth}
		\centering
		\fbox{
			\includegraphics[width=0.9\textwidth]{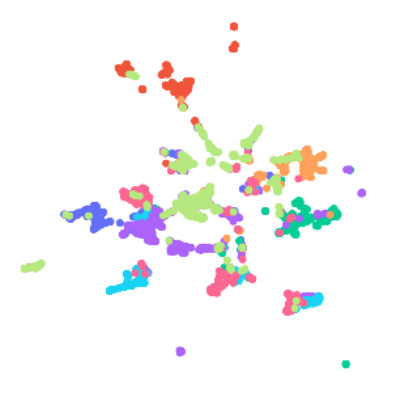}
		}
		\caption{Embeddings of entity tokens.}
		\label{fig:umap:ent}
	\end{subfigure}
	\begin{subfigure}[t]{0.33\textwidth}
		\centering
		\fbox{
			\includegraphics[width=0.9\textwidth]{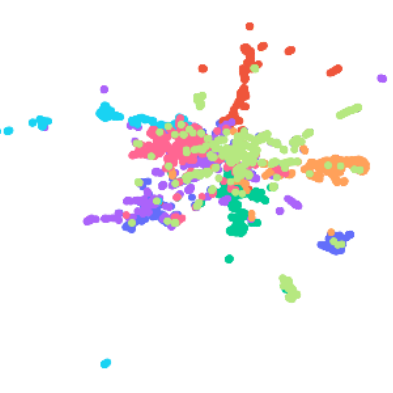}
		}
		\caption{Embeddings of entity mention tokens.}
		\label{fig:umap:wp}
	\end{subfigure}
	\begin{subfigure}[t]{0.3\textwidth}
		\centering
		\includegraphics[width=0.95\textwidth]{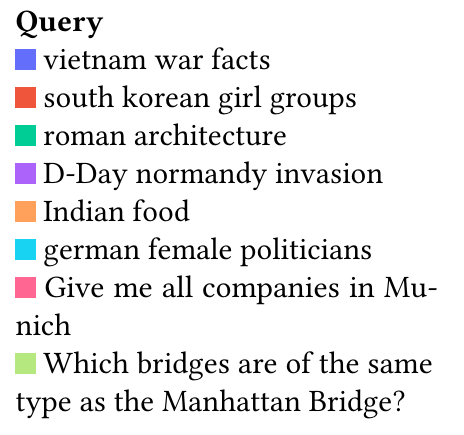}
	\end{subfigure}
	\miniskip
	\caption{UMAP plots of final layer EM-BERT embeddings of entity tokens and their corresponding mention tokens. Queries were randomly selected from queries with between 100 and 200 relevant entities. }
	\label{fig:umap}
\end{figure*}

\subsection{Query Analysis}
\label{sec:res:query}

In the following we answer our research question \emph{\textbf{RQ2:} When and which queries are helped by the EM-BERT model?}

\paragraph{\textbf{Entity annotations of queries}}
Knowing that not all queries in DBpedia-Entity v2 have a linked entity, we expect EM-BERT to perform better on queries with linked entities than on queries without them. 
To study this hypothesis, we compare NDCG@10 for two subsets of DBpedia-Entity v2 queries: with and without at least one linked entity.
Table~\ref{tab:difference_annotated_dbpedia} shows the results.
We see substantial improvements of EM-BERT over monoBERT and monoBERT (1st) for queries with linked entities, confirming our hypothesis.

\paragraph{\textbf{Tokenization of entity mentions}}
To further our understanding of helped queries, we investigate the difference in EM-BERT performance for various types of linked queries by looking into their mentions tokenization. The BERT tokenizer splits words into word pieces, which can be categorized into \emph{beginning} word pieces, indicating either the start of a word or a full word, and \emph{middle/end} word pieces starting with a `\#\#'. BERT employs around 30k tokens, consisting of around 24k beginning tokens and 6k middle/end tokens. 
Given that Wikipedia has around 12 million distinct entities, we expect only the most common entities to be present as a single word piece. We hypothesize that there is a performance difference between these very common entities, which are often seen in their natural form during the BERT pre-training process, and the rarer entities, where their mentions are broken into several word pieces.  We, therefore, divide entity mentions into four categories: 

\begin{itemize}[leftmargin=2em]
	\item \textbf{One token}: Mentions that are considered as a single token by the BERT tokenizer; e.g., ``France'' is tokenized to `France'.
	\item \textbf{Multiple tokens, no \#\#}: Mentions that are tokenized to more than one token, where each token represents a single unbroken word; e.g., ``Yoko Ono'' is tokenized to `Yoko', `Ono'.
	\item \textbf{One \#\#}: Mentions that are split into multiple word pieces, of which exactly one is a middle/end word piece, e.g., ``Weser'' is be tokenized to `wes', `\#\#er'
	\item \textbf{Multiple \#\#}: Mentions which gets split into multiple word pieces, of which more than one is a middle/end word piece, for example ``Frisian'' is tokenized to `fr', `\#\#isi', `\#\#an'.
\end{itemize}

We categorize each query with a linked entity in one of the previously mentioned categories, prioritizing the most complicated entity; i.e., if a query has two mentions belonging to  ``One token'' and ``Multiple \#\#'' categories, the query is categorized as ``Multiple \#\#''.
Figure~\ref{fig:tokenization} compares the performance of monoBERT and EM-BERT models for these four mention tokenization categories. It shows that gains of EM-BERT are more pronounced for queries in ``Multiple \#\#'' and ``One \#\#'' categories than other categories. This supports our claim that queries with uncommon entity mentions are most helped by including their entity embeddings in the EM-BERT model. 
Additionally, we illustrate the relation between different tokenization categories and the query types  mention (cf. Section~\ref{sec:exp:er}) in Figure~\ref{fig:heatmap}. 
The plot shows that 51\% of INEX-LD queries have no linked entities, explaining marginal improvements of EM-BERT for these queries. This plot also describes why EM-BERT performs best on ListSearch and QALD-2 queries: most of the less known entities fall in ``Multiple \#\#'' and ``One \#\#'' categories. 

These results provide an answer to RQ2: \emph{EM-BERT helps queries that are linked to at least one entity. Additionally,  entity information provided by the EM-BERT model is most useful for less popular entities, which BERT cannot treat their mentions as a single token.}

\begin{table*}[h]
\shrink
\caption{Comparison between EM-BERT and monoBERT for two example queries. Each query is listed twice, first with the normal BERT tokenization, then with the EM-BERT tokenization, both with the attentions of the first layer and first attention head. 
	The Rel. column indicates the relevance judgement for that entity.  The last two columns show the corresponding ranks obtained by the monoBERT/EM-BERT models for the entity.}
\shrink
\label{tab:examples_attention}
\begin{tabular}{ p{0.25\linewidth}  || p{0.40\linewidth} |p{0.03\linewidth} | p{0.12\linewidth} | p{0.1\linewidth} }
	\hline
	Query & Entity abstract & Rel.  & Baseline & Comparison \\
	\Xhline{2pt}
	{\setlength{\fboxsep}{0pt}\colorbox{white!0}{\parbox{\linewidth}{
				\colorbox{green!8.0}{\strut give} \colorbox{green!12.0}{\strut me} \colorbox{green!2.0}{\strut all} \colorbox{green!21.0}{\strut movies} \colorbox{green!6.0}{\strut directed} \colorbox{green!2.0}{\strut by} \colorbox{green!7.0}{\strut francis} \colorbox{green!11.0}{\strut ford} \colorbox{green!18.0}{\strut cop} \colorbox{green!28.0}{\strut \#\#pol} \colorbox{green!2.0}{\strut \#\#a} \colorbox{green!34.0}
	}}}& {\setlength{\fboxsep}{0pt}\colorbox{white!0}{\parbox{\linewidth}{
				\colorbox{green!16.0}{\strut christopher} \colorbox{green!17.0}{\strut cop} \colorbox{green!25.0}{\strut \#\#pol} \colorbox{green!2.0}{\strut \#\#a} \colorbox{green!9.0}{\strut (} \colorbox{green!7.0}{\strut born} \colorbox{green!6.0}{\strut january} \colorbox{green!0.0}{\strut 25} \colorbox{green!3.0}{\strut ,} \colorbox{green!13.0}{\strut 1962} \colorbox{green!7.0}{\strut )} \colorbox{green!2.0}{\strut is} \colorbox{green!4.0}{\strut a} \colorbox{green!9.0}{\strut film} \colorbox{green!4.0}{\strut director} \colorbox{green!4.0}{\strut and} \colorbox{green!7.0}{\strut producer} \colorbox{green!27.0} 
	}}} & 0 & monoBERT: 2 & EM-BERT: 551 \\
	\hline
	{\setlength{\fboxsep}{0pt}\colorbox{white!0}{\parbox{\linewidth}{\colorbox{green!6.0}{\strut give} \colorbox{green!11.0}{\strut me} \colorbox{green!2.0}{\strut all} \colorbox{green!18.0}{\strut movies} \colorbox{green!5.0}{\strut directed} \colorbox{green!2.0}{\strut by} \colorbox{green!6.0}{\strut francis} \colorbox{green!10.0}{\strut ford} \colorbox{green!15.0}{\strut cop} \colorbox{green!23.0}{\strut \#\#pol} \colorbox{green!2.0}{\strut \#\#a} \colorbox{green!4.0}{\strut /} \colorbox{green!39.0}{\strut ENTITY/Francis\_Ford\_Coppola} \colorbox{green!27.0}		
	}}} & 
	{\setlength{\fboxsep}{0pt}\colorbox{white!0}{\parbox{\linewidth}{
				\colorbox{green!14.0}{\strut rumble} \colorbox{green!14.0}{\strut fish} \colorbox{green!4.0}{\strut /} \colorbox{green!45.0}{\strut ENTITY/Rumble\_Fish} \colorbox{green!2.0}{\strut is} \colorbox{green!8.0}{\strut an} \colorbox{green!2.0}{\strut american} \colorbox{green!3.0}{\strut /} \colorbox{green!34.0}{\strut ENTITY/United\_States} \colorbox{green!17.0}{\strut 1983} \colorbox{green!9.0}{\strut drama} \colorbox{green!7.0}{\strut film} \colorbox{green!7.0}{\strut directed} \colorbox{green!1.0}{\strut by} \colorbox{green!4.0}{\strut francis} \colorbox{green!9.0}{\strut ford} \colorbox{green!10.0}{\strut cop} \colorbox{green!16.0}{\strut \#\#pol} \colorbox{green!2.0}{\strut \#\#a} \colorbox{green!3.0}{\strut /} \colorbox{green!30.0}{\strut ENTITY/Francis\_Ford\_Coppola} \colorbox{green!3.0}{\strut .} \colorbox{green!22.0}			
	}}} &2 & monoBERT: 119 & EM-BERT: 1 \\
	\hline \hline

	{\setlength{\fboxsep}{0pt}\colorbox{white!0}{\parbox{\linewidth}{
				\colorbox{green!11.0}{\strut give} \colorbox{green!16.0}{\strut me} \colorbox{green!8.0}{\strut the} \colorbox{green!8.0}{\strut capitals} \colorbox{green!5.0}{\strut of} \colorbox{green!2.0}{\strut all} \colorbox{green!4.0}{\strut countries} \colorbox{green!2.0}{\strut in} \colorbox{green!14.0}{\strut africa} \colorbox{green!50.0}					
				
	}}}  & 
	{\setlength{\fboxsep}{0pt}\colorbox{white!0}{\parbox{\linewidth}{
				\colorbox{green!5.0}{\strut list} \colorbox{green!6.0}{\strut of} \colorbox{green!6.0}{\strut african} \colorbox{green!30.0}{\strut depend} \colorbox{green!22.0}{\strut \#\#encies} \colorbox{green!23.0}{\strut ---} \colorbox{green!4.0}{\strut including} \colorbox{green!7.0}{\strut the} \colorbox{green!11.0}{\strut respective} \colorbox{green!4.0}{\strut capitals} \colorbox{green!5.0}{\strut .}				
				
	}}} & 0 & monoBERT: 1  & EM-BERT: 105 \\ 
	\hline
	{\setlength{\fboxsep}{0pt}\colorbox{white!0}{\parbox{\linewidth}{
				\colorbox{green!7.0}{\strut give} \colorbox{green!8.0}{\strut me} \colorbox{green!5.0}{\strut the} \colorbox{green!4.0}{\strut capitals} \colorbox{green!4.0}{\strut of} \colorbox{green!2.0}{\strut all} \colorbox{green!2.0}{\strut countries} \colorbox{green!2.0}{\strut in} \colorbox{green!7.0}{\strut africa} \colorbox{green!3.0}{\strut /} \colorbox{green!36.0}{\strut ENTITY/Africa}
				
	}}}  &
	{\setlength{\fboxsep}{0pt}\colorbox{white!0}{\parbox{\linewidth}{
				
				\colorbox{green!7.0}{\strut da} \colorbox{green!22.0}{\strut \#\#kar} \colorbox{green!4.0}{\strut /} \colorbox{green!28.0}{\strut ENTITY/Dakar} \colorbox{green!1.0}{\strut is} \colorbox{green!5.0}{\strut the} \colorbox{green!1.0}{\strut capital} \colorbox{green!2.0}{\strut and} \colorbox{green!12.0}{\strut largest} \colorbox{green!2.0}{\strut city} \colorbox{green!4.0}{\strut of} \colorbox{green!19.0}{\strut senegal} \colorbox{green!1.0}{\strut .} \colorbox{green!7.0}{\strut it} \colorbox{green!1.0}{\strut is} \colorbox{green!10.0}{\strut located} \colorbox{green!2.0}{\strut on} \colorbox{green!5.0}{\strut the} \colorbox{green!4.0}{\strut cap} \colorbox{green!3.0}{\strut -} \colorbox{green!5.0}{\strut ve} \colorbox{green!3.0}{\strut \#\#rt} \colorbox{green!5.0}{\strut peninsula} \colorbox{green!3.0}{\strut /} \colorbox{green!20.0}{\strut ENTITY/Cap-Vert} \colorbox{green!1.0}{\strut on} \colorbox{green!5.0}{\strut the} \colorbox{green!2.0}{\strut atlantic} \colorbox{green!1.0}{\strut coast} \colorbox{green!3.0}{\strut /} \colorbox{green!13.0}{\strut ENTITY/East\_Coast\_of\_the\_United\_States} \colorbox{green!2.0}{\strut and} \colorbox{green!1.0}{\strut is} \colorbox{green!4.0}{\strut the} \colorbox{green!1.0}{\strut western} \colorbox{green!28.0}{\strut \#\#most} \colorbox{green!1.0}{\strut city} \colorbox{green!1.0}{\strut in} \colorbox{green!5.0}{\strut the} \colorbox{green!3.0}{\strut old} \colorbox{green!1.0}{\strut world} \colorbox{green!3.0}{\strut /} \colorbox{green!42.0}{\strut ENTITY/Europe} \colorbox{green!2.0}{\strut and} \colorbox{green!1.0}{\strut on} \colorbox{green!5.0}{\strut the} \colorbox{green!1.0}{\strut african} \colorbox{green!3.0}{\strut /} \colorbox{green!25.0}{\strut ENTITY/Africa} \colorbox{green!6.0}{\strut mainland} \colorbox{green!2.0}{\strut .}
				
	}}}  &  1& monoBERT: 82 & EM-BERT: 1 \\ 

	\bottomrule

\end{tabular}
	\miniskip
\end{table*}

\subsection{Model Analysis}
\label{sec:res:model}

To investigate \emph{\textbf{RQ3:} Why does the EM-BERT model work and what does it learn during the fine-tuning stage?}, we examine the proximity of the embedding tokens compared to the mention tokens in the final layer of the BERT network, and also discuss several examples of helped queries.

\miniskip
\paragraph{\textbf{Entities vs. mention embeddings}}
To get an understanding of what BERT learns from entity information, we compare entity and mention embeddings in the final layer of EM-BERT. We randomly choose eight queries with 100-200 relevant entities. For each of those documents, we take the first four linked entities and extract their entity and mention embeddings from the final layer of EM-BERT. The mention embeddings are taken from the first starting token before that entity. We use UMAP~\cite{mcinnes:2018:UMAP} with the default settings to plot these embeddings in 2-dimensional space; see Figure~\ref{fig:umap}.
As is evident from the plots, the entity embeddings form more pronounced clusters than the entity mention embeddings.
For example, in the entity embedding plot, American bridges and rivers build a cluster together with other entities based in the US or Canada, while being close to the corresponding queries. This also indicates that the entity information injected through entity embeddings prevails into the final layers of the BERT model, keeping a better representation of entity information than provided with the default BERT word piece tokens.

\miniskip
\paragraph{\textbf{Insightful Examples}}
We discuss in detail a few (hand-picked) example queries that are especially helped or hurt by re-ranking with the EM-BERT model. Table~\ref{tab:examples_attention} visualizes the attention weights of the [CLS] token for the first attention head on the first layer, for sample query-entity pairs.
In the query ``Give me all movies directed by Francis Ford Coppola'', monoBERT incorrectly ranks other filmmaker members of the Coppola family, matching the last name only. EM-BERT, in contrast, finds the correct entity, as movies directed by Francis Ford Coppola are situated close to him in the embedding space (as opposed to the movies by his daughter Sofia Coppola). Table~\ref{tab:examples_attention} shows that EM-BERT gives high attention to the entity tokens relevant to the entity and query. It also shows high attention to the words ``movies'', which is essential here.

We can also see that through fine-tuning on DBpedia-entity, EM-BERT learns to retrieve specific entities for queries requesting a list. For example for the query 
``Give me the capitals of all countries in Africa'', while monoBERT's highest ranked entities are general overview pages like  `List of African dependencies', all of EM-BERT's highest-ranked entities are capitals of African countries like `Dakar' and `Porto-Novo'.


\subsection{Performance on Passage Retrieval} 
\label{sec:res:other}
Finally, we turn to our last research question, \emph{\textbf{RQ4:} How does our entity-enriched BERT-based model perform on other ranking tasks?}
To further our understanding on the effect of entity-enriched BERT for IR tasks, we compare the monoBERT and EM-BERT models after the first fine tuning stage on the MS MARCO collection. We observe no significant differences between the two models for both queries with and without linked entities. A similar observation is made for the TREC Conversation Assistant Track (TREC-CAsT)~\cite{Joko:2021:cast}.  We posit that the added value of enhancing BERT-based retrieval models with explicit entity representations is mainly pronounced for entity-oriented tasks.

%% file: conclusion.tex
\shrink
\section{Conclusion and future work}

In this research, we investigated the value of adding entity embeddings to BERT for entity search. We proposed an entity-enhanced BERT-based retrieval model, EM-BERT, and 
compared it with a regular BERT model. 
We found that EM-BERT improves substantially over the state-of-the-art results, showing that entity-oriented tasks benefit from entity embeddings in BERT, as is hypothesized in~\cite{broscheit:2019:investigating}.  
We found three categories where entities help the most: complex natural language queries, queries requesting a list of entities with a certain property, and queries with less known entities, in which their mentions get split in multiple tokens by the BERT tokenizer.
We also showed that EM-BERT, unlike BERT-based retrieval models, can be fine-tuned with limited training data, thus being effective for lesser-resourced entity-related tasks. 
For further research, the effect of the entity linking method can be investigated; here we used REL, which is an entity linker with high precision. 
It would be also interesting to evaluate the performance of other dense and sparse retrieval methods on entity-oriented queries and tasks. This, however, may require additional training data for fine-tuning on entity-oriented queries, 
which would create a supplementary MS Marco subset to train and evaluate ranking systems. 
\balance

%% file: sigir22.bbl

\begin{thebibliography}{54}


\ifx \showCODEN    \undefined \def \showCODEN     #1{\unskip}     \fi
\ifx \showDOI      \undefined \def \showDOI       #1{#1}\fi
\ifx \showISBNx    \undefined \def \showISBNx     #1{\unskip}     \fi
\ifx \showISBNxiii \undefined \def \showISBNxiii  #1{\unskip}     \fi
\ifx \showISSN     \undefined \def \showISSN      #1{\unskip}     \fi
\ifx \showLCCN     \undefined \def \showLCCN      #1{\unskip}     \fi
\ifx \shownote     \undefined \def \shownote      #1{#1}          \fi
\ifx \showarticletitle \undefined \def \showarticletitle #1{#1}   \fi
\ifx \showURL      \undefined \def \showURL       {\relax}        \fi
\providecommand\bibfield[2]{#2}
\providecommand\bibinfo[2]{#2}
\providecommand\natexlab[1]{#1}
\providecommand\showeprint[2][]{arXiv:#2}

\bibitem[\protect\citeauthoryear{Akbik, Blythe, and Vollgraf}{Akbik
  et~al\mbox{.}}{2018}]%
        {akbik:2018:contextual}
\bibfield{author}{\bibinfo{person}{Alan Akbik}, \bibinfo{person}{Duncan
  Blythe}, {and} \bibinfo{person}{Roland Vollgraf}.}
  \bibinfo{year}{2018}\natexlab{}.
\newblock \showarticletitle{Contextual String Embeddings for Sequence
  Labeling}. In \bibinfo{booktitle}{\emph{Proc. of 27th International
  Conference on Computational Linguistics}} \emph{(\bibinfo{series}{COLING
  '18})}. \bibinfo{pages}{1638--1649}.
\newblock


\bibitem[\protect\citeauthoryear{Balog}{Balog}{2018}]%
        {Balog:2018:EOS}
\bibfield{author}{\bibinfo{person}{Krisztian Balog}.}
  \bibinfo{year}{2018}\natexlab{}.
\newblock \bibinfo{booktitle}{\emph{{Entity-Oriented Search}}}.
  \bibinfo{series}{The Information Retrieval Series},
  Vol.~\bibinfo{volume}{39}.
\newblock \bibinfo{publisher}{Springer}.
\newblock
\showISBNx{978-3-319-93933-9}


\bibitem[\protect\citeauthoryear{Blanco, Ottaviano, and Meij}{Blanco
  et~al\mbox{.}}{2015}]%
        {Blanco:2015:FSE}
\bibfield{author}{\bibinfo{person}{Roi Blanco}, \bibinfo{person}{Giuseppe
  Ottaviano}, {and} \bibinfo{person}{Edgar Meij}.}
  \bibinfo{year}{2015}\natexlab{}.
\newblock \showarticletitle{{Fast and Space-Efficient Entity Linking in
  Queries}}.
\newblock \bibinfo{journal}{\emph{Proc. of the Eighth ACM International
  Conference on Web Search and Data Mining}} (\bibinfo{year}{2015}),
  \bibinfo{pages}{179--188}.
\newblock


\bibitem[\protect\citeauthoryear{Bordes, Usunier, Garcia-Duran, Weston, and
  Yakhnenko}{Bordes et~al\mbox{.}}{2013}]%
        {bordes:2013:translating}
\bibfield{author}{\bibinfo{person}{Antoine Bordes}, \bibinfo{person}{Nicolas
  Usunier}, \bibinfo{person}{Alberto Garcia-Duran}, \bibinfo{person}{Jason
  Weston}, {and} \bibinfo{person}{Oksana Yakhnenko}.}
  \bibinfo{year}{2013}\natexlab{}.
\newblock \showarticletitle{Translating embeddings for modeling
  multi-relational data}. In \bibinfo{booktitle}{\emph{Proc. of the 26th
  International Conference on Neural Information Processing Systems}}
  \emph{(\bibinfo{series}{NeurIPS '13})}. \bibinfo{pages}{2787--2795}.
\newblock


\bibitem[\protect\citeauthoryear{Broscheit}{Broscheit}{2019}]%
        {broscheit:2019:investigating}
\bibfield{author}{\bibinfo{person}{Samuel Broscheit}.}
  \bibinfo{year}{2019}\natexlab{}.
\newblock \showarticletitle{Investigating Entity Knowledge in {BERT} with
  Simple Neural End-To-End Entity Linking}. In \bibinfo{booktitle}{\emph{Proc.
  of the 23rd Conference on Computational Natural Language Learning}}
  \emph{(\bibinfo{series}{CoNLL '19})}. \bibinfo{pages}{677--685}.
\newblock


\bibitem[\protect\citeauthoryear{Cao, Izacard, Riedel, and Petroni}{Cao
  et~al\mbox{.}}{2021}]%
        {decao:2020:autoregressive}
\bibfield{author}{\bibinfo{person}{Nicola~De Cao}, \bibinfo{person}{Gautier
  Izacard}, \bibinfo{person}{Sebastian Riedel}, {and} \bibinfo{person}{Fabio
  Petroni}.} \bibinfo{year}{2021}\natexlab{}.
\newblock \showarticletitle{Autoregressive Entity Retrieval}. In
  \bibinfo{booktitle}{\emph{Proc. of the International Conference on Learning
  Representations}} \emph{(\bibinfo{series}{ICLR '21})}.
\newblock


\bibitem[\protect\citeauthoryear{Cornolti, Ferragina, Ciaramita, R\"{u}d, and
  Sch\"{u}tze}{Cornolti et~al\mbox{.}}{2016}]%
        {cornolti:2016:piggyback}
\bibfield{author}{\bibinfo{person}{Marco Cornolti}, \bibinfo{person}{Paolo
  Ferragina}, \bibinfo{person}{Massimiliano Ciaramita}, \bibinfo{person}{Stefan
  R\"{u}d}, {and} \bibinfo{person}{Hinrich Sch\"{u}tze}.}
  \bibinfo{year}{2016}\natexlab{}.
\newblock \showarticletitle{A Piggyback System for Joint Entity Mention
  Detection and Linking in Web Queries}. In \bibinfo{booktitle}{\emph{Proc. of
  the 25th International Conference on World Wide Web}}
  \emph{(\bibinfo{series}{WWW '16})}. \bibinfo{pages}{567--578}.
\newblock


\bibitem[\protect\citeauthoryear{Dalton, ~, and Callan}{Dalton
  et~al\mbox{.}}{2020}]%
        {dalton:2020:trec}
\bibfield{author}{\bibinfo{person}{Jeffrey Dalton}, \bibinfo{person}{Chenyan
  }, {and} \bibinfo{person}{Jamie Callan}.} \bibinfo{year}{2020}\natexlab{}.
\newblock \showarticletitle{TREC CAsT 2019: The conversational assistance track
  overview}.
\newblock \bibinfo{journal}{\emph{arXiv preprint}} (\bibinfo{year}{2020}).
\newblock
\showeprint{2003.13624}


\bibitem[\protect\citeauthoryear{Dalton, Dietz, and Allan}{Dalton
  et~al\mbox{.}}{2014}]%
        {Dalton:2014:EQF}
\bibfield{author}{\bibinfo{person}{Jeffrey Dalton}, \bibinfo{person}{Laura
  Dietz}, {and} \bibinfo{person}{James Allan}.}
  \bibinfo{year}{2014}\natexlab{}.
\newblock \showarticletitle{{Entity Query Feature Expansion Using Knowledge
  Base Links}}. In \bibinfo{booktitle}{\emph{Proc. of the 37th International
  ACM SIGIR Conference on Research and Development in Information Retrieval}}
  \emph{(\bibinfo{series}{SIGIR '14})}. \bibinfo{pages}{365--374}.
\newblock


\bibitem[\protect\citeauthoryear{Daza, Cochez, and Groth}{Daza
  et~al\mbox{.}}{2021}]%
        {daza:2021:inductive}
\bibfield{author}{\bibinfo{person}{Daniel Daza}, \bibinfo{person}{Michael
  Cochez}, {and} \bibinfo{person}{Paul Groth}.}
  \bibinfo{year}{2021}\natexlab{}.
\newblock \showarticletitle{Inductive Entity Representations from Text via Link
  Prediction}. In \bibinfo{booktitle}{\emph{Proc. of the Web Conference 2021}}
  \emph{(\bibinfo{series}{WWW '21})}. \bibinfo{pages}{798–808}.
\newblock


\bibitem[\protect\citeauthoryear{Devlin, Chang, Lee, and Toutanova}{Devlin
  et~al\mbox{.}}{2019}]%
        {devlin:2019:bert}
\bibfield{author}{\bibinfo{person}{Jacob Devlin}, \bibinfo{person}{Ming-Wei
  Chang}, \bibinfo{person}{Kenton Lee}, {and} \bibinfo{person}{Kristina
  Toutanova}.} \bibinfo{year}{2019}\natexlab{}.
\newblock \showarticletitle{{BERT}: Pre-training of Deep Bidirectional
  Transformers for Language Understanding}. In \bibinfo{booktitle}{\emph{Proc.
  of The North American Chapter of the Association for Computational
  Linguistics '19}} \emph{(\bibinfo{series}{NAACL '19})}.
  \bibinfo{pages}{4171--4186}.
\newblock


\bibitem[\protect\citeauthoryear{Dietz}{Dietz}{2019}]%
        {dietz:2019:ENT}
\bibfield{author}{\bibinfo{person}{Laura Dietz}.}
  \bibinfo{year}{2019}\natexlab{}.
\newblock \showarticletitle{ENT Rank: Retrieving Entities for Topical
  Information Needs through Entity-Neighbor-Text Relations}. In
  \bibinfo{booktitle}{\emph{Proc. of the 42nd International ACM SIGIR
  Conference on Research and Development in Information Retrieval}}
  \emph{(\bibinfo{series}{SIGIR'19})}. \bibinfo{pages}{215–224}.
\newblock


\bibitem[\protect\citeauthoryear{Dodge, Ilharco, Schwartz, Farhadi, Hajishirzi,
  and Smith}{Dodge et~al\mbox{.}}{2020}]%
        {Dodge:2020:FPL}
\bibfield{author}{\bibinfo{person}{Jesse Dodge}, \bibinfo{person}{Gabriel
  Ilharco}, \bibinfo{person}{Roy Schwartz}, \bibinfo{person}{Ali Farhadi},
  \bibinfo{person}{Hannaneh Hajishirzi}, {and} \bibinfo{person}{Noah Smith}.}
  \bibinfo{year}{2020}\natexlab{}.
\newblock \showarticletitle{Fine-tuning pretrained language models: Weight
  initializations, data orders, and early stopping}.
\newblock \bibinfo{journal}{\emph{arXiv preprint arXiv:2002.06305}}
  (\bibinfo{year}{2020}).
\newblock


\bibitem[\protect\citeauthoryear{Ferragina and Scaiella}{Ferragina and
  Scaiella}{2010}]%
        {ferragina:2010:TAGME}
\bibfield{author}{\bibinfo{person}{Paolo Ferragina} {and} \bibinfo{person}{Ugo
  Scaiella}.} \bibinfo{year}{2010}\natexlab{}.
\newblock \showarticletitle{TAGME: On-the-Fly Annotation of Short Text
  Fragments (by Wikipedia Entities)}. In \bibinfo{booktitle}{\emph{Proc. of the
  International Conference on Information and Knowledge Management}}
  \emph{(\bibinfo{series}{CIKM '10})}. \bibinfo{pages}{1625--1628}.
\newblock


\bibitem[\protect\citeauthoryear{Garigliotti, Hasibi, and Balog}{Garigliotti
  et~al\mbox{.}}{2019}]%
        {Garigliotti:2019:IET}
\bibfield{author}{\bibinfo{person}{Dar{\'{i}}o Garigliotti},
  \bibinfo{person}{Faegheh Hasibi}, {and} \bibinfo{person}{Krisztian Balog}.}
  \bibinfo{year}{2019}\natexlab{}.
\newblock \showarticletitle{{Identifying and Exploiting Target Entity Type
  Information for Ad hoc Entity Retrieval}}.
\newblock \bibinfo{journal}{\emph{Information Retrieval Journal}}
  \bibinfo{volume}{22}, \bibinfo{number}{3} (\bibinfo{year}{2019}),
  \bibinfo{pages}{285--323}.
\newblock
\showISSN{1573-7659}


\bibitem[\protect\citeauthoryear{Gerritse, Hasibi, and De~Vries}{Gerritse
  et~al\mbox{.}}{2020}]%
        {gerritse:2020:graph}
\bibfield{author}{\bibinfo{person}{Emma Gerritse}, \bibinfo{person}{Faegheh
  Hasibi}, {and} \bibinfo{person}{Arjen De~Vries}.}
  \bibinfo{year}{2020}\natexlab{}.
\newblock \showarticletitle{Graph-Embedding Empowered Entity Retrieval}. In
  \bibinfo{booktitle}{\emph{Proc. of European Conference on Information
  Retrieval}} \emph{(\bibinfo{series}{ECIR '20})}.
\newblock


\bibitem[\protect\citeauthoryear{Hasibi, Balog, and Bratsberg}{Hasibi
  et~al\mbox{.}}{2015}]%
        {Hasibi:2015:ELQ}
\bibfield{author}{\bibinfo{person}{Faegheh Hasibi}, \bibinfo{person}{Krisztian
  Balog}, {and} \bibinfo{person}{Svein~Erik Bratsberg}.}
  \bibinfo{year}{2015}\natexlab{}.
\newblock \showarticletitle{{Entity Linking in Queries: Tasks and Evaluation}}.
  In \bibinfo{booktitle}{\emph{Proc. of the 2015 International Conference on
  The Theory of Information Retrieval}} \emph{(\bibinfo{series}{ICTIR '15})}.
  \bibinfo{pages}{171--180}.
\newblock


\bibitem[\protect\citeauthoryear{Hasibi, Balog, and Bratsberg}{Hasibi
  et~al\mbox{.}}{2016}]%
        {Hasibi:2016:EEL}
\bibfield{author}{\bibinfo{person}{Faegheh Hasibi}, \bibinfo{person}{Krisztian
  Balog}, {and} \bibinfo{person}{Svein~Erik Bratsberg}.}
  \bibinfo{year}{2016}\natexlab{}.
\newblock \showarticletitle{{Exploiting Entity Linking in Queries for Entity
  Retrieval}}. In \bibinfo{booktitle}{\emph{Proc. of the 2016 ACM International
  Conference on the Theory of Information Retrieval}}
  \emph{(\bibinfo{series}{ICTIR '16})}. \bibinfo{pages}{209--218}.
\newblock


\bibitem[\protect\citeauthoryear{Hasibi, Balog, and Bratsberg}{Hasibi
  et~al\mbox{.}}{2017a}]%
        {Hasibi:2017:ELQ}
\bibfield{author}{\bibinfo{person}{Faegheh Hasibi}, \bibinfo{person}{Krisztian
  Balog}, {and} \bibinfo{person}{Svein~Erik Bratsberg}.}
  \bibinfo{year}{2017}\natexlab{a}.
\newblock \showarticletitle{Entity Linking in Queries: Efficiency vs.
  Effectiveness}. In \bibinfo{booktitle}{\emph{Proc. of the 39th European
  Conference on Information Retrieval}} \emph{(\bibinfo{series}{ECIR '17})}.
  \bibinfo{pages}{40--53}.
\newblock


\bibitem[\protect\citeauthoryear{Hasibi, Balog, Garigliotti, and Zhang}{Hasibi
  et~al\mbox{.}}{2017b}]%
        {Hasibi:2017:NTE}
\bibfield{author}{\bibinfo{person}{Faegheh Hasibi}, \bibinfo{person}{Krisztian
  Balog}, \bibinfo{person}{Daŕio Garigliotti}, {and} \bibinfo{person}{Shuo
  Zhang}.} \bibinfo{year}{2017}\natexlab{b}.
\newblock \showarticletitle{{Nordlys: A toolkit for entity-oriented and
  semantic search}}. In \bibinfo{booktitle}{\emph{Proc. of the 40th
  International ACM SIGIR Conference on Research and Development in Information
  Retrieval}} \emph{(\bibinfo{series}{SIGIR '17})}.
  \bibinfo{pages}{1289--1292}.
\newblock


\bibitem[\protect\citeauthoryear{Hasibi, Nikolaev, Xiong, Balog, Bratsberg,
  Kotov, and Callan}{Hasibi et~al\mbox{.}}{2017c}]%
        {hasibi:2017:dbpedia}
\bibfield{author}{\bibinfo{person}{Faegheh Hasibi}, \bibinfo{person}{Fedor
  Nikolaev}, \bibinfo{person}{Chenyan Xiong}, \bibinfo{person}{Krisztian
  Balog}, \bibinfo{person}{Svein~Erik Bratsberg}, \bibinfo{person}{Alexander
  Kotov}, {and} \bibinfo{person}{Jamie Callan}.}
  \bibinfo{year}{2017}\natexlab{c}.
\newblock \showarticletitle{{DBpedia-Entity V2}: A Test Collection for Entity
  Search}. In \bibinfo{booktitle}{\emph{Proc. of the 40th International ACM
  SIGIR Conference on Research and Development in Information Retrieval}}
  \emph{(\bibinfo{series}{SIGIR '19})}. \bibinfo{pages}{1265--1268}.
\newblock


\bibitem[\protect\citeauthoryear{Hofst{\"a}tter, Althammer, Schr{\"o}der,
  Sertkan, and Hanbury}{Hofst{\"a}tter et~al\mbox{.}}{2020}]%
        {hofstaetter:2020:crossarchitecturekd}
\bibfield{author}{\bibinfo{person}{Sebastian Hofst{\"a}tter},
  \bibinfo{person}{Sophia Althammer}, \bibinfo{person}{Michael Schr{\"o}der},
  \bibinfo{person}{Mete Sertkan}, {and} \bibinfo{person}{Allan Hanbury}.}
  \bibinfo{year}{2020}\natexlab{}.
\newblock \showarticletitle{Improving Efficient Neural Ranking Models with
  Cross-Architecture Knowledge Distillation}.
\newblock \bibinfo{journal}{\emph{arXiv preprint}} (\bibinfo{year}{2020}).
\newblock
\showeprint{2010.02666}


\bibitem[\protect\citeauthoryear{Jiang, Anastasopoulos, Araki, Ding, and
  Neubig}{Jiang et~al\mbox{.}}{2020}]%
        {Jiang:2020:XMF}
\bibfield{author}{\bibinfo{person}{Zhengbao Jiang}, \bibinfo{person}{Antonios
  Anastasopoulos}, \bibinfo{person}{Jun Araki}, \bibinfo{person}{Haibo Ding},
  {and} \bibinfo{person}{Graham Neubig}.} \bibinfo{year}{2020}\natexlab{}.
\newblock \showarticletitle{{X}-{FACTR}: Multilingual Factual Knowledge
  Retrieval from Pretrained Language Models}. In
  \bibinfo{booktitle}{\emph{Proc. of the 2020 Conference on Empirical Methods
  in Natural Language Processing}} \emph{(\bibinfo{series}{EMNLP '20})}.
  \bibinfo{pages}{5943--5959}.
\newblock


\bibitem[\protect\citeauthoryear{Joko, Gerritse, Hasibi, and de~Vries}{Joko
  et~al\mbox{.}}{2022}]%
        {Joko:2021:cast}
\bibfield{author}{\bibinfo{person}{Hideaki Joko}, \bibinfo{person}{Emma~J
  Gerritse}, \bibinfo{person}{Faegheh Hasibi}, {and} \bibinfo{person}{Arjen~P
  de Vries}.} \bibinfo{year}{2022}\natexlab{}.
\newblock \showarticletitle{Radboud University at TREC CAsT 2021}. In
  \bibinfo{booktitle}{\emph{TREC}}.
\newblock


\bibitem[\protect\citeauthoryear{Joshi, Chen, Liu, Weld, Zettlemoyer, and
  Levy}{Joshi et~al\mbox{.}}{2020}]%
        {joshi:2019:spanbert}
\bibfield{author}{\bibinfo{person}{Mandar Joshi}, \bibinfo{person}{Danqi Chen},
  \bibinfo{person}{Yinhan Liu}, \bibinfo{person}{Daniel~S. Weld},
  \bibinfo{person}{Luke Zettlemoyer}, {and} \bibinfo{person}{Omer Levy}.}
  \bibinfo{year}{2020}\natexlab{}.
\newblock \showarticletitle{{S}pan{BERT}: Improving Pre-training by
  Representing and Predicting Spans}.
\newblock \bibinfo{journal}{\emph{Transactions of the Association for
  Computational Linguistics}} (\bibinfo{year}{2020}), \bibinfo{pages}{64--77}.
\newblock


\bibitem[\protect\citeauthoryear{Joshi, Levy, Zettlemoyer, and Weld}{Joshi
  et~al\mbox{.}}{2019}]%
        {joshi:2019:bert}
\bibfield{author}{\bibinfo{person}{Mandar Joshi}, \bibinfo{person}{Omer Levy},
  \bibinfo{person}{Luke Zettlemoyer}, {and} \bibinfo{person}{Daniel Weld}.}
  \bibinfo{year}{2019}\natexlab{}.
\newblock \showarticletitle{{{BERT} for Coreference Resolution: Baselines and
  Analysis}}. In \bibinfo{booktitle}{\emph{Proc. of the 2019 Conference on
  Empirical Methods in Natural Language Processing and the 9th International
  Joint Conference on Natural Language Processing}}
  \emph{(\bibinfo{series}{EMNLP-IJCNLP '19})}. \bibinfo{pages}{5803--5808}.
\newblock


\bibitem[\protect\citeauthoryear{Liu, Ott, Goyal, Du, Joshi, Chen, Levy, Lewis,
  Zettlemoyer, and Stoyanov}{Liu et~al\mbox{.}}{2019}]%
        {liu:2020:roberta}
\bibfield{author}{\bibinfo{person}{Yinhan Liu}, \bibinfo{person}{Myle Ott},
  \bibinfo{person}{Naman Goyal}, \bibinfo{person}{Jingfei Du},
  \bibinfo{person}{Mandar Joshi}, \bibinfo{person}{Danqi Chen},
  \bibinfo{person}{Omer Levy}, \bibinfo{person}{Mike Lewis},
  \bibinfo{person}{Luke Zettlemoyer}, {and} \bibinfo{person}{Veselin
  Stoyanov}.} \bibinfo{year}{2019}\natexlab{}.
\newblock \showarticletitle{RoBERTa: A Robustly Optimized BERT Pretraining
  Approach}.
\newblock \bibinfo{journal}{\emph{arXiv preprint}} (\bibinfo{year}{2019}).
\newblock
\showeprint{1907.11692}


\bibitem[\protect\citeauthoryear{McInnes, Healy, Saul, and Grossberger}{McInnes
  et~al\mbox{.}}{2018}]%
        {mcinnes:2018:UMAP}
\bibfield{author}{\bibinfo{person}{Leland McInnes}, \bibinfo{person}{John
  Healy}, \bibinfo{person}{Nathaniel Saul}, {and} \bibinfo{person}{Lukas
  Grossberger}.} \bibinfo{year}{2018}\natexlab{}.
\newblock \showarticletitle{UMAP: Uniform Manifold Approximation and
  Projection}.
\newblock \bibinfo{journal}{\emph{The Journal of Open Source Software}}
  \bibinfo{volume}{3}, \bibinfo{number}{29} (\bibinfo{year}{2018}),
  \bibinfo{pages}{861}.
\newblock


\bibitem[\protect\citeauthoryear{Mikolov, Sutskever, Chen, Corrado, and
  Dean}{Mikolov et~al\mbox{.}}{2013}]%
        {mikolov:2013:distributed}
\bibfield{author}{\bibinfo{person}{Tomas Mikolov}, \bibinfo{person}{Ilya
  Sutskever}, \bibinfo{person}{Kai Chen}, \bibinfo{person}{Gregory~S. Corrado},
  {and} \bibinfo{person}{Jeffrey Dean}.} \bibinfo{year}{2013}\natexlab{}.
\newblock \showarticletitle{Distributed Representations of Words and Phrases
  and their Compositionality}. In \bibinfo{booktitle}{\emph{Proc. of Advances
  in Neural Information Processing Systems}} \emph{(\bibinfo{series}{NeurIPS
  '13})}. \bibinfo{pages}{3111--3119}.
\newblock


\bibitem[\protect\citeauthoryear{Nguyen, Rosenberg, Song, Gao, Tiwary,
  Majumder, and Deng}{Nguyen et~al\mbox{.}}{2016}]%
        {nguyen:2016:msmarco}
\bibfield{author}{\bibinfo{person}{Tri Nguyen}, \bibinfo{person}{Mir
  Rosenberg}, \bibinfo{person}{Xia Song}, \bibinfo{person}{Jianfeng Gao},
  \bibinfo{person}{Saurabh Tiwary}, \bibinfo{person}{Rangan Majumder}, {and}
  \bibinfo{person}{Li Deng}.} \bibinfo{year}{2016}\natexlab{}.
\newblock \showarticletitle{{MS} {MARCO:} {A} Human Generated MAchine Reading
  COmprehension Dataset}. In \bibinfo{booktitle}{\emph{Proc. of the Workshop on
  Cognitive Computation: Integrating neural and symbolic approaches 2016}}.
\newblock


\bibitem[\protect\citeauthoryear{Nikolaev and Kotov}{Nikolaev and
  Kotov}{2020}]%
        {nikolaev:2020:joint}
\bibfield{author}{\bibinfo{person}{Fedor Nikolaev} {and}
  \bibinfo{person}{Alexander Kotov}.} \bibinfo{year}{2020}\natexlab{}.
\newblock \showarticletitle{Joint Word and Entity Embeddings for Entity
  Retrieval from a Knowledge Graph}. In \bibinfo{booktitle}{\emph{Proc. of the
  European Conference on Information Retrieval}} \emph{(\bibinfo{series}{ECIR
  '20})}. \bibinfo{pages}{141--155}.
\newblock


\bibitem[\protect\citeauthoryear{Nogueira and Cho}{Nogueira and Cho}{2019}]%
        {nogueira:2019:passage}
\bibfield{author}{\bibinfo{person}{Rodrigo Nogueira} {and}
  \bibinfo{person}{Kyunghyun Cho}.} \bibinfo{year}{2019}\natexlab{}.
\newblock \showarticletitle{Passage Re-ranking with BERT}.
\newblock \bibinfo{journal}{\emph{arXiv preprint}} (\bibinfo{year}{2019}).
\newblock
\showeprint{1901.04085}


\bibitem[\protect\citeauthoryear{Nogueira, Yang, Cho, and Lin}{Nogueira
  et~al\mbox{.}}{2019}]%
        {nogueira:2019:multi}
\bibfield{author}{\bibinfo{person}{Rodrigo Nogueira}, \bibinfo{person}{Wei
  Yang}, \bibinfo{person}{Kyunghyun Cho}, {and} \bibinfo{person}{Jimmy Lin}.}
  \bibinfo{year}{2019}\natexlab{}.
\newblock \showarticletitle{Multi-stage document ranking with BERT}.
\newblock \bibinfo{journal}{\emph{arXiv preprint}} (\bibinfo{year}{2019}).
\newblock
\showeprint{1910.14424}


\bibitem[\protect\citeauthoryear{Peters, Neumann, Logan, Schwartz, Joshi,
  Singh, and Smith}{Peters et~al\mbox{.}}{2019}]%
        {peters:2019:knowledge}
\bibfield{author}{\bibinfo{person}{Matthew~E. Peters}, \bibinfo{person}{Mark
  Neumann}, \bibinfo{person}{Robert Logan}, \bibinfo{person}{Roy Schwartz},
  \bibinfo{person}{Vidur Joshi}, \bibinfo{person}{Sameer Singh}, {and}
  \bibinfo{person}{Noah~A. Smith}.} \bibinfo{year}{2019}\natexlab{}.
\newblock \showarticletitle{Knowledge Enhanced Contextual Word
  Representations}. In \bibinfo{booktitle}{\emph{Proc. of the 2019 Conference
  on Empirical Methods in Natural Language Processing and the 9th International
  Joint Conference on Natural Language Processing}}
  \emph{(\bibinfo{series}{EMNLP-IJCNLP '19})}. \bibinfo{pages}{43--54}.
\newblock


\bibitem[\protect\citeauthoryear{Petroni, Rockt{\"{a}}schel, Lewis, Bakhtin,
  Wu, Miller, and Riedel}{Petroni et~al\mbox{.}}{2019}]%
        {Petroni:2019:LMK}
\bibfield{author}{\bibinfo{person}{Fabio Petroni}, \bibinfo{person}{Tim
  Rockt{\"{a}}schel}, \bibinfo{person}{Patrick S.~H. Lewis},
  \bibinfo{person}{Anton Bakhtin}, \bibinfo{person}{Yuxiang Wu},
  \bibinfo{person}{Alexander~H. Miller}, {and} \bibinfo{person}{Sebastian
  Riedel}.} \bibinfo{year}{2019}\natexlab{}.
\newblock \showarticletitle{Language Models as Knowledge Bases?}. In
  \bibinfo{booktitle}{\emph{Proc. of the 2019 Conference on Empirical Methods
  in Natural Language Processing and the 9th International Joint Conference on
  Natural Language Processing}} \emph{(\bibinfo{series}{EMNLP-IJCNLP '19})}.
  \bibinfo{pages}{2463--2473}.
\newblock


\bibitem[\protect\citeauthoryear{Phang, Fevry, and Bowman}{Phang
  et~al\mbox{.}}{2019}]%
        {Phang:2019:STILT}
\bibfield{author}{\bibinfo{person}{Jason Phang}, \bibinfo{person}{Thibault
  Fevry}, {and} \bibinfo{person}{Samuel~R. Bowman}.}
  \bibinfo{year}{2019}\natexlab{}.
\newblock \showarticletitle{Sentence Encoders on STILTs: Supplementary Training
  on Intermediate Labeled-data Tasks}.
\newblock \bibinfo{journal}{\emph{arXiv preprint arXiv:1811.01088}}
  (\bibinfo{year}{2019}).
\newblock


\bibitem[\protect\citeauthoryear{Poerner, Waltinger, and Sch{\"u}tze}{Poerner
  et~al\mbox{.}}{2020}]%
        {poerner:2020:ebert}
\bibfield{author}{\bibinfo{person}{Nina Poerner}, \bibinfo{person}{Ulli
  Waltinger}, {and} \bibinfo{person}{Hinrich Sch{\"u}tze}.}
  \bibinfo{year}{2020}\natexlab{}.
\newblock \showarticletitle{{E}-{BERT}: Efficient-Yet-Effective Entity
  Embeddings for {BERT}}. In \bibinfo{booktitle}{\emph{Findings of the
  Association for Computational Linguistics}} \emph{(\bibinfo{series}{ELMNLP
  '20})}. \bibinfo{pages}{803--818}.
\newblock


\bibitem[\protect\citeauthoryear{Raffel, Shazeer, Roberts, Lee, Narang, Matena,
  Zhou, Li, and Liu}{Raffel et~al\mbox{.}}{2020}]%
        {raffel:2020:exploring}
\bibfield{author}{\bibinfo{person}{Colin Raffel}, \bibinfo{person}{Noam
  Shazeer}, \bibinfo{person}{Adam Roberts}, \bibinfo{person}{Katherine Lee},
  \bibinfo{person}{Sharan Narang}, \bibinfo{person}{Michael Matena},
  \bibinfo{person}{Yanqi Zhou}, \bibinfo{person}{Wei Li}, {and}
  \bibinfo{person}{Peter~J. Liu}.} \bibinfo{year}{2020}\natexlab{}.
\newblock \showarticletitle{Exploring the Limits of Transfer Learning with a
  Unified Text-to-Text Transformer}.
\newblock \bibinfo{journal}{\emph{Journal of Machine Learning Research}}
  \bibinfo{volume}{21}, \bibinfo{number}{140} (\bibinfo{year}{2020}),
  \bibinfo{pages}{1--67}.
\newblock


\bibitem[\protect\citeauthoryear{Talmor, Elazar, Goldberg, and Berant}{Talmor
  et~al\mbox{.}}{2020}]%
        {Talmor:2020:oLMpics}
\bibfield{author}{\bibinfo{person}{Alon Talmor}, \bibinfo{person}{Yanai
  Elazar}, \bibinfo{person}{Yoav Goldberg}, {and} \bibinfo{person}{Jonathan
  Berant}.} \bibinfo{year}{2020}\natexlab{}.
\newblock \showarticletitle{oLMpics - On what Language Model Pre-training
  Captures}.
\newblock \bibinfo{journal}{\emph{Transactions of the Association for
  Computational Linguistics}}  \bibinfo{volume}{8} (\bibinfo{year}{2020}),
  \bibinfo{pages}{743--758}.
\newblock


\bibitem[\protect\citeauthoryear{Thakur, Reimers, R{\"u}ckl{\'e}, Srivastava,
  and Gurevych}{Thakur et~al\mbox{.}}{2021}]%
        {thakur:2021:beir}
\bibfield{author}{\bibinfo{person}{Nandan Thakur}, \bibinfo{person}{Nils
  Reimers}, \bibinfo{person}{Andreas R{\"u}ckl{\'e}}, \bibinfo{person}{Abhishek
  Srivastava}, {and} \bibinfo{person}{Iryna Gurevych}.}
  \bibinfo{year}{2021}\natexlab{}.
\newblock \showarticletitle{{BEIR}: A Heterogeneous Benchmark for Zero-shot
  Evaluation of Information Retrieval Models}. In
  \bibinfo{booktitle}{\emph{Proc. of 35th Conference on Neural Information
  Processing Systems Datasets and Benchmarks Track (Round 2)}}
  \emph{(\bibinfo{series}{NeurIPS '21})}.
\newblock


\bibitem[\protect\citeauthoryear{van Hulst, Hasibi, Dercksen, Balog, and
  de~Vries}{van Hulst et~al\mbox{.}}{2020}]%
        {vanHulst:2020:rel}
\bibfield{author}{\bibinfo{person}{Johannes~M. van Hulst},
  \bibinfo{person}{Faegheh Hasibi}, \bibinfo{person}{Koen Dercksen},
  \bibinfo{person}{Krisztian Balog}, {and} \bibinfo{person}{Arjen~P. de
  Vries}.} \bibinfo{year}{2020}\natexlab{}.
\newblock \showarticletitle{REL: An Entity Linker Standing on the Shoulders of
  Giants}. In \bibinfo{booktitle}{\emph{Proc. of the 43rd International ACM
  SIGIR Conference on Research and Development in Information Retrieval}}
  \emph{(\bibinfo{series}{SIGIR '20})}.
\newblock


\bibitem[\protect\citeauthoryear{Wang, Liu, and Song}{Wang
  et~al\mbox{.}}{2020}]%
        {wang:2020:language}
\bibfield{author}{\bibinfo{person}{Chenguang Wang}, \bibinfo{person}{Xiao Liu},
  {and} \bibinfo{person}{Dawn Song}.} \bibinfo{year}{2020}\natexlab{}.
\newblock \showarticletitle{Language Models are Open Knowledge Graphs}.
\newblock \bibinfo{journal}{\emph{arXiv preprint}} (\bibinfo{year}{2020}).
\newblock
\showeprint{2010.11967}


\bibitem[\protect\citeauthoryear{{Wang}, {Tang}, {Duan}, {Wei}, {Huang}, {ji},
  {Cao}, {Jiang}, and {Zhou}}{{Wang} et~al\mbox{.}}{2020}]%
        {wang:2020:kadapter}
\bibfield{author}{\bibinfo{person}{Ruize {Wang}}, \bibinfo{person}{Duyu
  {Tang}}, \bibinfo{person}{Nan {Duan}}, \bibinfo{person}{Zhongyu {Wei}},
  \bibinfo{person}{Xuanjing {Huang}}, \bibinfo{person}{Jianshu {ji}},
  \bibinfo{person}{Guihong {Cao}}, \bibinfo{person}{Daxin {Jiang}}, {and}
  \bibinfo{person}{Ming {Zhou}}.} \bibinfo{year}{2020}\natexlab{}.
\newblock \showarticletitle{{K-Adapter: Infusing Knowledge into Pre-Trained
  Models with Adapters}}.
\newblock \bibinfo{journal}{\emph{arXiv preprint}} (\bibinfo{year}{2020}).
\newblock
\showeprint{2002.01808}


\bibitem[\protect\citeauthoryear{Wang, Wei, Dong, Bao, Yang, and Zhou}{Wang
  et~al\mbox{.}}{2020}]%
        {wang:2020:minilm}
\bibfield{author}{\bibinfo{person}{Wenhui Wang}, \bibinfo{person}{Furu Wei},
  \bibinfo{person}{Li Dong}, \bibinfo{person}{Hangbo Bao}, \bibinfo{person}{Nan
  Yang}, {and} \bibinfo{person}{Ming Zhou}.} \bibinfo{year}{2020}\natexlab{}.
\newblock \showarticletitle{MiniLM: Deep Self-Attention Distillation for
  Task-Agnostic Compression of Pre-Trained Transformers}. In
  \bibinfo{booktitle}{\emph{Proc. of Advances in Neural Information Processing
  Systems}} \emph{(\bibinfo{series}{NeurIPS '21})},
  \bibfield{editor}{\bibinfo{person}{H.~Larochelle},
  \bibinfo{person}{M.~Ranzato}, \bibinfo{person}{R.~Hadsell},
  \bibinfo{person}{M.~F. Balcan}, {and} \bibinfo{person}{H.~Lin}} (Eds.).
  \bibinfo{pages}{5776--5788}.
\newblock


\bibitem[\protect\citeauthoryear{Wang, Gao, Zhu, Liu, Li, and Tang}{Wang
  et~al\mbox{.}}{2019a}]%
        {wang:2019:kepler}
\bibfield{author}{\bibinfo{person}{Xiaozhi Wang}, \bibinfo{person}{Tianyu Gao},
  \bibinfo{person}{Zhaocheng Zhu}, \bibinfo{person}{Zhiyuan Liu},
  \bibinfo{person}{Juanzi Li}, {and} \bibinfo{person}{Jian Tang}.}
  \bibinfo{year}{2019}\natexlab{a}.
\newblock \showarticletitle{KEPLER: A unified model for knowledge embedding and
  pre-trained language representation}.
\newblock \bibinfo{journal}{\emph{arXiv preprint}} (\bibinfo{year}{2019}).
\newblock
\showeprint{1911.06136}


\bibitem[\protect\citeauthoryear{Wang, Ng, Ma, Nallapati, and Xiang}{Wang
  et~al\mbox{.}}{2019b}]%
        {wang:2019:multi}
\bibfield{author}{\bibinfo{person}{Zhiguo Wang}, \bibinfo{person}{Patrick Ng},
  \bibinfo{person}{Xiaofei Ma}, \bibinfo{person}{Ramesh Nallapati}, {and}
  \bibinfo{person}{Bing Xiang}.} \bibinfo{year}{2019}\natexlab{b}.
\newblock \showarticletitle{Multi-passage {BERT}: A Globally Normalized {BERT}
  Model for Open-domain Question Answering}. In \bibinfo{booktitle}{\emph{Proc.
  of the 2019 Conference on Empirical Methods in Natural Language Processing
  and the 9th International Joint Conference on Natural Language Processing}}
  \emph{(\bibinfo{series}{EMNLP-IJCNLP '19})}. \bibinfo{pages}{5878--5882}.
\newblock


\bibitem[\protect\citeauthoryear{Xiong, Callan, and Liu}{Xiong
  et~al\mbox{.}}{2017a}]%
        {Xiong:2017:WED}
\bibfield{author}{\bibinfo{person}{Chenyan Xiong}, \bibinfo{person}{Jamie
  Callan}, {and} \bibinfo{person}{Tie-Yan Liu}.}
  \bibinfo{year}{2017}\natexlab{a}.
\newblock \showarticletitle{{Word-Entity Duet Representations for Document
  Ranking}}. In \bibinfo{booktitle}{\emph{Proc. of the 40th International ACM
  SIGIR Conference on Research and Development in Information Retrieval}}
  \emph{(\bibinfo{series}{SIGIR '17})}. \bibinfo{pages}{763--772}.
\newblock


\bibitem[\protect\citeauthoryear{Xiong, Power, and Callan}{Xiong
  et~al\mbox{.}}{2017b}]%
        {Xiong:2017:ESR}
\bibfield{author}{\bibinfo{person}{Chenyan Xiong}, \bibinfo{person}{Russell
  Power}, {and} \bibinfo{person}{Jamie Callan}.}
  \bibinfo{year}{2017}\natexlab{b}.
\newblock \showarticletitle{{Explicit Semantic Ranking for Academic Search via
  Knowledge Graph Embedding}}. In \bibinfo{booktitle}{\emph{Proc. of the 26th
  International Conference on World Wide Web}} \emph{(\bibinfo{series}{WWW
  '17})}. \bibinfo{pages}{1271--1279}.
\newblock


\bibitem[\protect\citeauthoryear{Yamada, Asai, Sakuma, Shindo, Takeda,
  Takefuji, and Matsumoto}{Yamada et~al\mbox{.}}{2020a}]%
        {yamada:2020:wikipedia2vec}
\bibfield{author}{\bibinfo{person}{Ikuya Yamada}, \bibinfo{person}{Akari Asai},
  \bibinfo{person}{Jin Sakuma}, \bibinfo{person}{Hiroyuki Shindo},
  \bibinfo{person}{Hideaki Takeda}, \bibinfo{person}{Yoshiyasu Takefuji}, {and}
  \bibinfo{person}{Yuji Matsumoto}.} \bibinfo{year}{2020}\natexlab{a}.
\newblock \showarticletitle{{W}ikipedia2{V}ec: An Efficient Toolkit for
  Learning and Visualizing the Embeddings of Words and Entities from
  {W}ikipedia}. In \bibinfo{booktitle}{\emph{Proc. of the 2020 Conference on
  Empirical Methods in Natural Language Processing: System Demonstrations}}
  \emph{(\bibinfo{series}{EMNLP '20})}. \bibinfo{pages}{23--30}.
\newblock


\bibitem[\protect\citeauthoryear{Yamada, Asai, Shindo, Takeda, and
  Matsumoto}{Yamada et~al\mbox{.}}{2020b}]%
        {yamada:2020:luke}
\bibfield{author}{\bibinfo{person}{Ikuya Yamada}, \bibinfo{person}{Akari Asai},
  \bibinfo{person}{Hiroyuki Shindo}, \bibinfo{person}{Hideaki Takeda}, {and}
  \bibinfo{person}{Yuji Matsumoto}.} \bibinfo{year}{2020}\natexlab{b}.
\newblock \showarticletitle{LUKE: Deep Contextualized Entity Representations
  with Entity-aware Self-attention}. In \bibinfo{booktitle}{\emph{Proc. of the
  2020 Conference on Empirical Methods in Natural Language Processing}}
  \emph{(\bibinfo{series}{EMNLP '20})}. \bibinfo{pages}{6442--6454}.
\newblock


\bibitem[\protect\citeauthoryear{Yamada, Shindo, Takeda, and Takefuji}{Yamada
  et~al\mbox{.}}{2016}]%
        {yamada:2016:joint}
\bibfield{author}{\bibinfo{person}{Ikuya Yamada}, \bibinfo{person}{Hiroyuki
  Shindo}, \bibinfo{person}{Hideaki Takeda}, {and} \bibinfo{person}{Yoshiyasu
  Takefuji}.} \bibinfo{year}{2016}\natexlab{}.
\newblock \showarticletitle{Joint learning of the embedding of words and
  entities for named entity disambiguation}. In \bibinfo{booktitle}{\emph{Proc.
  of The SIGNLL Conference on Computational Natural Language Learning}}
  \emph{(\bibinfo{series}{SIGNLL '16})}. \bibinfo{pages}{250--259}.
\newblock


\bibitem[\protect\citeauthoryear{Yang, Dai, Yang, Carbonell, Salakhutdinov, and
  Le}{Yang et~al\mbox{.}}{2019}]%
        {yang:2019:XLNet}
\bibfield{author}{\bibinfo{person}{Zhilin Yang}, \bibinfo{person}{Zihang Dai},
  \bibinfo{person}{Yiming Yang}, \bibinfo{person}{Jaime Carbonell},
  \bibinfo{person}{Russ~R Salakhutdinov}, {and} \bibinfo{person}{Quoc~V Le}.}
  \bibinfo{year}{2019}\natexlab{}.
\newblock \showarticletitle{XLNet: Generalized Autoregressive Pretraining for
  Language Understanding}. In \bibinfo{booktitle}{\emph{Advances in Neural
  Information Processing Systems}} \emph{(\bibinfo{series}{NeurIPS '19})}.
  \bibinfo{pages}{5753–5763}.
\newblock


\bibitem[\protect\citeauthoryear{Zhang, Wu, Katiyar, Weinberger, and
  Artzi}{Zhang et~al\mbox{.}}{2021}]%
        {zhang:2020:revisiting}
\bibfield{author}{\bibinfo{person}{Tianyi Zhang}, \bibinfo{person}{Felix Wu},
  \bibinfo{person}{Arzoo Katiyar}, \bibinfo{person}{Kilian~Q Weinberger}, {and}
  \bibinfo{person}{Yoav Artzi}.} \bibinfo{year}{2021}\natexlab{}.
\newblock \showarticletitle{Revisiting Few-sample BERT Fine-tuning}. In
  \bibinfo{booktitle}{\emph{Proc. of 2021 International Conference on Learning
  Representations}} \emph{(\bibinfo{series}{ICLR '20})}.
\newblock


\bibitem[\protect\citeauthoryear{Zhang, Han, Liu, Jiang, Sun, and Liu}{Zhang
  et~al\mbox{.}}{2019}]%
        {zhang:2019:ernie}
\bibfield{author}{\bibinfo{person}{Zhengyan Zhang}, \bibinfo{person}{Xu Han},
  \bibinfo{person}{Zhiyuan Liu}, \bibinfo{person}{Xin Jiang},
  \bibinfo{person}{Maosong Sun}, {and} \bibinfo{person}{Qun Liu}.}
  \bibinfo{year}{2019}\natexlab{}.
\newblock \showarticletitle{{ERNIE}: Enhanced Language Representation with
  Informative Entities}. In \bibinfo{booktitle}{\emph{Proc. of the 57th Annual
  Meeting of the Association for Computational Linguistics}}
  \emph{(\bibinfo{series}{ACL '19})}. \bibinfo{pages}{1441--1451}.
\newblock


\end{thebibliography}
